\documentclass[prl,superscriptaddress,twocolumn,amsmath]{revtex4-1}
\usepackage{graphicx}
\usepackage{color}
\usepackage{multirow}
\usepackage{hhline}
\usepackage{siunitx}
\usepackage{color}
\usepackage{draftwatermark}
\SetWatermarkText{}
\SetWatermarkScale{1}
\usepackage[usenames,dvipsnames]{xcolor}

\begin{document}

\title{Deterministic Single-Phonon Source Triggered by a Single Photon}

\author{Immo S\"{o}llner} \email{immo.soellner@unibas.ch; now at the University of Basel }
\affiliation{Niels Bohr Institute, University of Copenhagen, Blegdamsvej 17, DK-2100 Copenhagen, Denmark}
%\affiliation{Department of Physics, University of Basel, Klingelbergstrasse 82, CH-4056 Basel, Switzerland}
\author{Leonardo Midolo}
\affiliation{Niels Bohr Institute, University of Copenhagen, Blegdamsvej 17, DK-2100 Copenhagen, Denmark}
\author{Peter Lodahl}
\affiliation{Niels Bohr Institute, University of Copenhagen, Blegdamsvej 17, DK-2100 Copenhagen, Denmark}

\date{\today}

\begin{abstract}
  We propose a scheme that enables the deterministic generation of single phonons at GHz frequencies triggered by single photons in the near infrared. This process is mediated by a quantum dot embedded on-chip in an opto-mechanical circuit, which allows for the simultaneous control of the relevant photonic and phononic frequencies. We devise new opto-mechanical circuit elements that constitute the necessary building blocks for the proposed scheme and are readily implementable within the current state-of-the-art of nano-fabrication. This will open new avenues for implementing quantum functionalities based on phonons as an on-chip quantum bus.
\end{abstract}

\maketitle
%\begin{counted}
Engineering of periodic nanostructures has proven to be an immensely powerful tool to shape the properties of a material. Most notably are photonic crystals, which are created by the periodic modulation of the refractive index. Their photonic bandgaps and defect modes have been widely studied \cite{Joannopoulos2008} and have found a number of applications in nano-photonics. For example, in solid-state quantum optics, they are commonly used to control light--matter interaction \cite{Lodahl2013}. Similarly, the periodic modulation of mechanical properties leads to the formation of phononic crystals \cite{Gorishnyy2005a}. Chip-scale devices that allow engineering of both the photonic and the phononic density of states simultaneously have recently been proposed \cite{Maldovan2006,Eichenfield2009,Safavi-naeini2010a,Mohammadi2010}. This has lead to a range of theoretical \cite{Safavi-Naeini2011,Galland2014,Schmidt2015} and experimental \cite{Chan2011,Cohen2015,Riedinger2015} breakthroughs in opto-mechanics. In parallel there has been significant progress on coupling single solid-state emitters, in the form of nitrogen-vacancy centers or self assembled quantum dots(QDs), to mechanical resonators either via magnetic gradient coupling \cite{Rabl2009,Arcizet2011} or strain coupling \cite{Wilson-Rae2004, Bennett2013, Ovartchaiyapong2014, Barfuss2015, MacQuarrie2013, Yeo2013,Montinaro2014}. Additionally, the potential of using phononic crystals to control single-phonon mediated processes in solid-state systems has recently been alluded to \cite{Colless2014,Jahnke2015}.

In optics, the reliable generation and detection of single-photon and entangled-photon states has important applications within quantum information science. Similar goals are being pursued in opto-mechanics where the generation \cite{Galland2014,Flayac2014,Riedinger2015} and detection \cite{Cohen2015} of non-classical phononic states are opening new avenues of research. These schemes are typically probabilistic and rely on the direct radiation-pressure interaction between co-localized optical and mechanical modes where the coupling rate between the modes is proportional to the square root of the intra-cavity photon number \cite{Aspelmeyer2014}. However, parasitic absorption is a problem for large intra-cavity photon numbers when operating at milikelvin temperatures \cite{Meenehan2014,Riedinger2015}, which becomes necessary when performing experiments on single phonons with frequencies in the few GHz regime. In this paper we propose an alternative approach based on a hybrid opto-mechanical(OM)-crystal where we engineer the coupling of a three-level emitter to both the photonic and the phononic reservoirs. We demonstrate how the internal spin state of the emitter can be used to mediate strong photon-phonon interaction for an emitter embedded in the OM-crystal. In contrast to the standard approach in opto-mechanics \cite{Galland2014,Flayac2014,Cohen2015,Aspelmeyer2014}, our proposal does not rely on radiation-pressure coupling, which is strong for large intra-cavity photon numbers. Instead the deterministic single-photon--single-phonon cascade triggered by a single narrow-bandwidth photon is operated at an average intra-cavity photon numbers significantly below one \footnote{\label{footnote1}We refer to Ref.~\cite{Sollner2015} for a discussion about what constitutes narrow bandwidth for the incident photon and the role of dephasing in the coherent scattering regime.}, thus offering a route to circumvent the problem of parasitic heating. We propose a readily implementable device that can be used with QD emitters.

\begin{figure}[t!]
	\includegraphics{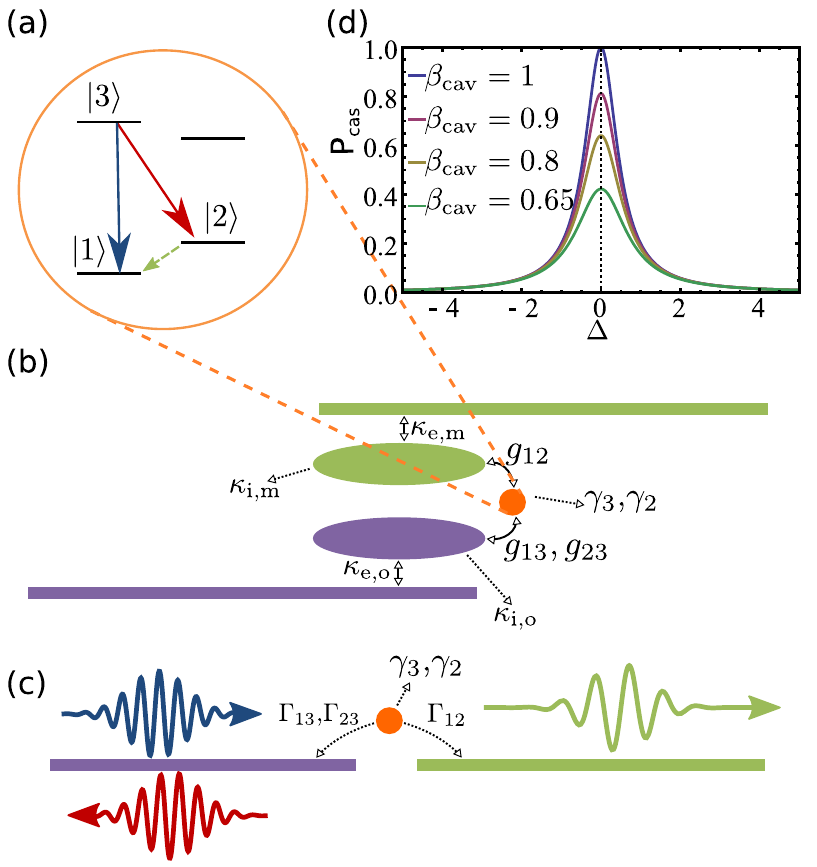}
\caption{ \label{fig:fig1} Operational principle of the deterministic single-photon--single-phonon cascade. {\bf a)} Level structure of a singly charged exciton in a magnetic field in Voigt-configuration. The optical transitions between states $|1\rangle$ and $|3\rangle$ and states $|2\rangle$ and $|3\rangle$ form a lambda system, indicated by the solid arrows. Furthermore, the two ground-states, $|1\rangle$  and $|2\rangle$, are  coupled to each other via a single-phonon-mediated transition, indicated by the dashed arrow. {\bf b)} Schematic of the desired OM-circuit functioning as a heralded single-phonon source, which is deterministically triggered by a single photon. The photonic (phononic) waveguide, shown in purple on the bottom (green at the top), couples the single photons (phonons) in-to and out-of the OM-cavity modes (middle) at the rate $\kappa_{\textrm{e,o}}$ ($\kappa_{\textrm{e,m}}$). The QD couples to the photonic and the co-localized phononic mode of the OM-cavity with the rates $g_{13},g_{23}$, and $g_{12}$ and to the loss modes in the photonic and phononic environments with rates $\gamma_{3}$ and $\gamma_{2}$. {\bf c)} The schematic of the effective OM-circuit for the parameter regime discussed in the text and with the relevant rates indicated. The incident photon (blue wavepackets) triggers the outgoing photon-phonon cascade (red wavepacket and green wavepacket). {\bf d)} The success probability of initializing the photon--phonon pair by a single photon incident on the trigger transition as a function of detuning of the trigger photon, $\Delta=\omega_{\textrm{tr}}-\omega_{13}$, for several values of the cavity-$\beta$-factors, $\beta_{\textrm{cav}}=(\Gamma_{13}+\Gamma_{23})/(\Gamma_{13}+\Gamma_{23}+\gamma_3)$. }
\end{figure}

Our protocol is based on the two optical transitions of a lambda-system, given by a singly-charged QD (trion) in an in-plane magnetic field (Voigt-configuration), c.f. Fig.~\ref{fig:fig1}(a). In this configuration there are two allowed linearly-polarized optical transitions that decay at the same rate \cite{Press2008}, while the two optical ground states are coupled by a spin-flip rate. Experiments have shown that it is possible to operate in a regime (depending on the Zeeman-splitting, sample temperature, and the co-tunnelling rates) where the spin-flip process is dominated by single-acoustic-phonon mediated transitions \cite{Kroutvar2004,Dreiser2008}, which is in good agreement with theoretical predictions \cite{Khaetskii2001,Woods2002,Golovach2004}. We note that the coherence properties of the emitted phonons are inherited from the coherence of the spin. In many QD experiments the spin coherence is not limited by phonon mediated relaxation processes \cite{Warburton2013}, leading to emission of incoherent phonons. However, recently progress has been made towards increasing the spin-coherence times \cite{Prechtel2015} and the ultimately limiting process remains to be determined, especially when operating at mK-temperatures. The emitter is embedded in an OM-circuit shown schematically in Fig.~\ref{fig:fig1}(b). A photonic (phononic) waveguide couples the single photons (phonons) in-to and out-of the OM-circuit. The photonic and phononic waveguides each couple to their respective mode of the OM-cavity at the rates denoted $\kappa_{\textrm{e,o}}$ and $\kappa_{\textrm{e,m}}$. All relevant rates are indicated in Fig.~\ref{fig:fig1}(b) and controlling their relative magnitudes offer a wide range of design possibilities. For the remainder of this paper we will focus on just one of these different realizations, which is well suited for the deterministic generation of single phonons. We consider the case where both cavity modes are in the over-coupled regime, meaning that the resonator loss is dominated by the coupling to the waveguide mode, $\kappa_{\textrm{e,o}}$($\kappa_{\textrm{e,m}}$)$\gg\kappa_{\textrm{i,o}}$($\kappa_{\textrm{i,m}}$) \cite{Galland2014}. Thus the intrinsic cavity loss rates, $\kappa_{\textrm{i,o}}$ and $\kappa_{\textrm{i,m}}$, can be neglected in the following. The emitter--resonator coupling is in the bad cavity but large cooperativity regime, i.e.,  $\kappa_{\textrm{e,o}}\gg g_{\textrm{13}},g_{\textrm{23}}$ and $\kappa_{\textrm{e,m}}\gg g_{\textrm{12}}$ but $4(g_{\textrm{13}}^2+g_{\textrm{23}}^2)/(\kappa_{\textrm{e,o}}\gamma_{3})\gg1$ and $2g_{\textrm{12}}^2/(\kappa_{\textrm{e,m}}\gamma_{2})\gg1$ \cite{Rice1988}. In the bad cavity regime the influence of the cavity on an emitter close to resonance can be captured by a cavity-enhanced effective decay rate of the excited state into the cavity mode: $\Gamma_{12}=2g_{\textrm{12}}^2/\kappa_{\textrm{e,m}}$, $\Gamma_{13}=2g_{\textrm{13}}^2/\kappa_{\textrm{e,o}}$, and $\Gamma_{23}=2g_{\textrm{23}}^2/\kappa_{\textrm{e,o}}$ \cite{Rice1988}. Thus the schematic in Fig.~\ref{fig:fig1}(b) simplifies to the effective circuit shown in Fig.~\ref{fig:fig1}(c) for the implementation we are considering here. This is best described as an emitter coupled to unidirectional photonic and phononic reservoirs \cite{Rephaeli2013,Ralph2015}. The single-frequency scattering coefficient for the elastic scattering of a trigger photon, $\omega_{\textrm{tr}}$, incident on the $|3\rangle \rightarrow |1\rangle$ transition is given by
\begin{equation}
t_{\textrm{tr}} = {{\Delta-i(\Gamma_{13}-\Gamma_{23}-\gamma_3)/2}\over{\Delta+i(\Gamma_{13}+\Gamma_{23}+\gamma_3)/2}}.%t_{\textrm{tr}} = {{(\omega_{\textrm{tr}}- \omega_{13})-i(\Gamma_{13}-\Gamma_{23}-\gamma_3)/2}\over{(\omega_{\textrm{tr}}- \omega_{13})+i(\Gamma_{13}+\Gamma_{23}+\gamma_3)/2}}.
\end{equation}
For the case of $\Gamma_{13}=\Gamma_{23}+\gamma_3$ and $\Delta=0$, i.e., when the two optical transitions are enhanced by the same amount and the incident photon is on resonance and is narrow-band compared to the QD transition, we see complete destructive interference of the scattered light at the resonance frequency $\omega_{13}$. Thus, the photon has to be scattered along one of the remaining available decay channels \cite{Bradford2012,Shomroni2014}. Consequently, the scattering coefficient for a trigger photon incident on the $|3\rangle \rightarrow |1\rangle$ transition to Raman-scatter along the $|3\rangle \rightarrow |2\rangle$ transition is,
\begin{equation}\label{trigcas_scatt}
t_{\textrm{cas}} = {{-i\sqrt{\Gamma_{13}\Gamma_{23}}}\over{\Delta+i(\Gamma_{13}+\Gamma_{23}+\gamma_3)/2}}.%t_{\textrm{tr}} = {{(\omega_{\textrm{tr}}- \omega_{13})-i(\Gamma_{13}-\Gamma_{23}-\gamma_3)/2}\over{(\omega_{\textrm{tr}}- \omega_{13})+i(\Gamma_{13}+\Gamma_{23}+\gamma_3)/2}}.
\end{equation}
From Eq.(~\ref{trigcas_scatt}) we calculate the success probability of initializing the single photon -- single phonon cascade process, c.f. Fig.~\ref{fig:fig1}(d), which approaches unity for large coupling-efficiencies of the optical transitions. It is promising to note that recent experiments in photonic-crystal waveguides have shown that even for moderate enhancements of the waveguide mode, spontaneous-emission coupling-efficiencies $\beta_{\textrm{wg}} >98\%$ can be readily achieved \cite{Arcari2014}. This is largely the result of the strongly suppressed coupling to optical loss modes for QDs embedded in  photonic-crystal membranes \cite{Wang2011}. The incident and outgoing wavepackets for the successful generation of a photon--phonon cascade from a single incident trigger photon is illustrated in Fig.~\ref{fig:fig1}(c).

\begin{figure*}[ht!]
	\includegraphics{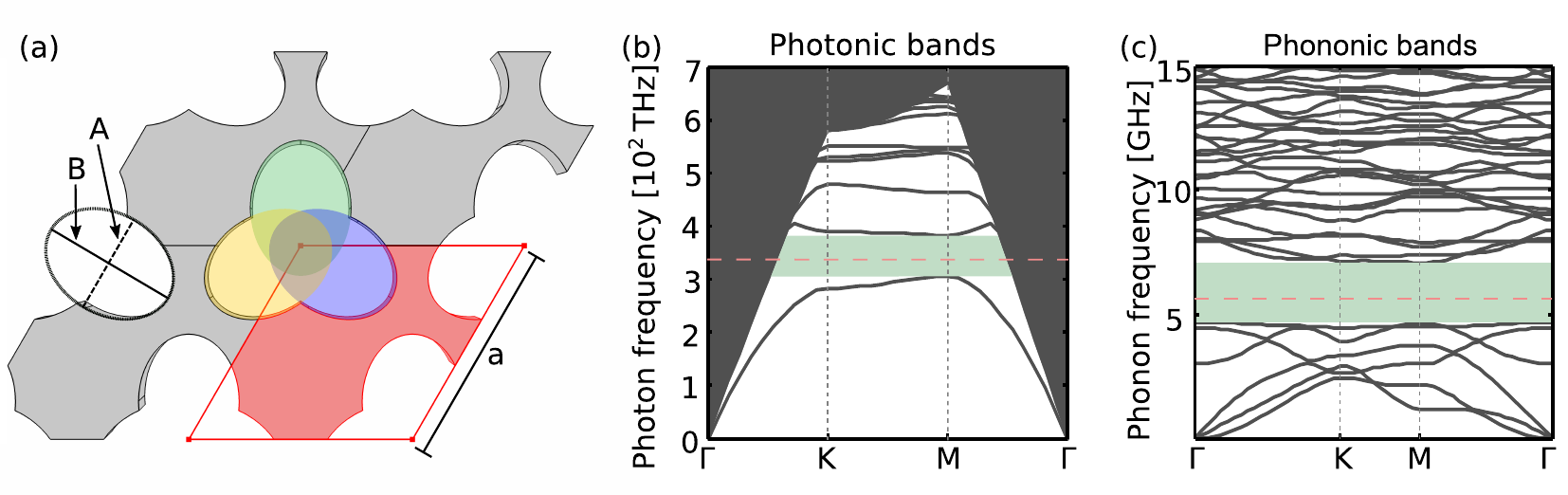}
\caption{ \label{fig:fig2} Design of the OM-crystal membrane. {\bf a)} Illustration of the OM-crystal with a single unit-cell highlighted. The holes in the membrane consist of three overlapping ellipsoids, indicated by the three shaded regions, rotated by $2\pi/3$ with respect to one another. The lattice constant of the crystal is denoted by a and the minor and major axis of the ellipsoids are denoted by A and B. {\bf b)} The photonic band-structure of the shamrock-crystal with parameters (a, A, B, L, d)=(300nm, 0.45a, 0.6a, 0.17a, 0.65a). The symmetries and the irreducible Brillouin zone of this crystal are discussed in the SM. For the parameters used here we find a band gap for the TE-like slab modes in the frequency range $\SI{305}{\tera\hertz}$ to $\SI{383}{\tera\hertz}$. {\bf c)} Phononic band-structure for the same crystal parameters displaying a complete phononic bandgap in the range $\SI{4.7}{\giga\hertz}$ to $\SI{7.1}{\giga\hertz}$. The horizontal dashed lines in {\bf b)} and {\bf c)} indicate the photonic and phononic resonance frequencies of the cavity modes shown in Fig.~\ref{fig:fig1}(b). The photonic resonance frequency was chosen to fall within the range of optical transitions of standard InGaAs QDs and the phononic resonance frequency corresponds to a Zeeman energy-splitting  achievable in standard cryomagnet systems. }
\end{figure*}
Implementing the circuit in Fig.~\ref{fig:fig1}(b) requires an OM-crystal that supports photonic and phononic band gaps at frequencies relevant to the optical transitions~\cite{Lodahl2013} and the Zeeman-splitting~\cite{Warburton2013} obtainable for the considered emitter, i.e., InGaAs QDs. The OM-crystal we propose (shown in Fig.~\ref{fig:fig2}(a)) consists of a hexagonal array of holes, separated by the lattice constant $\textrm{a}=\SI{300}{\nano\meter}$ etched into a GaAs membrane of thickness d=0.65a. Each hole is made up by three overlapping ellipses rotated by $2\pi/3$ with respect to one another. The minor and major axes of the ellipse are given by A=0.45a and B=0.6a, respectively. The center of each ellipse is shifted outwards along its major axis by L=0.17a resulting in a final shape that is reminiscent of a shamrock, c.f. Fig.~\ref{fig:fig2}(a). This leads to a reduction of the crystal symmetry compared to conventional photonic \cite{Lodahl2013} and previous OM-crystals \cite{Safavi-naeini2010a} (c.f. the Supplementary Material (SM) for more information \footnote{\label{footnote2}See Supplemental Material at [URL will be inserted by publisher] for more information on the reduction of the crystal symmetry.}). A similar structure has been investigate for its photonic properties \cite{Wen2008}.

The photonic band-diagram for modes with TE-like symmetry is shown in Fig.~\ref{fig:fig2}(b). The lightly shaded region in the center indicates the in-plane band-gap and the dashed line marks the position of the optical transitions within the band-gap. The dark shaded region at the sides and the top of Fig.~\ref{fig:fig2}(b) indicates the continuum of leaky radiation modes, i.e., modes that are not confined to the membrane. For simulations of the phononic bands (see SM for details) we consider the modes of all symmetries, c.f. Fig.~\ref{fig:fig2}(c), taking into account the anisotropy of the elastic constants of gallium arsenide. In the phononic band structure there are no leaky modes, as all modes are confined to the membrane. Thus, contrary to the photonic case, the phononic band-diagrams exhibit a complete band-gap. This has promising implications for the control of single-phonon mediated transitions as they can be completely suppressed within the band-gap, where the phononic density of states drops to zero. For our proposal this means that the coupling efficiency of the $|2\rangle \rightarrow |1\rangle$ transition to the phononic cavity mode is only limited by the probability to decay through a single-phonon process, as opposed to other spin-relaxation processes such as co-tunneling \cite{Dreiser2008} and multi-phonon effects \cite{Khaetskii2001,Woods2002,Trif2009}. This differs from the coupling efficiencies in photonic systems, where in addition to the probability of the emitter decaying through a single photon process, the coupling efficiency is limited by the coupling to non-guided radiation modes \cite{Arcari2014}. Nonetheless, completely analogous to optical emitters in photonic crystals, the coupling rate to the target mode can be  enhanced by increasing the local density of mechanical states through the use of defect modes.

\begin{figure*}[ht!]
	\includegraphics{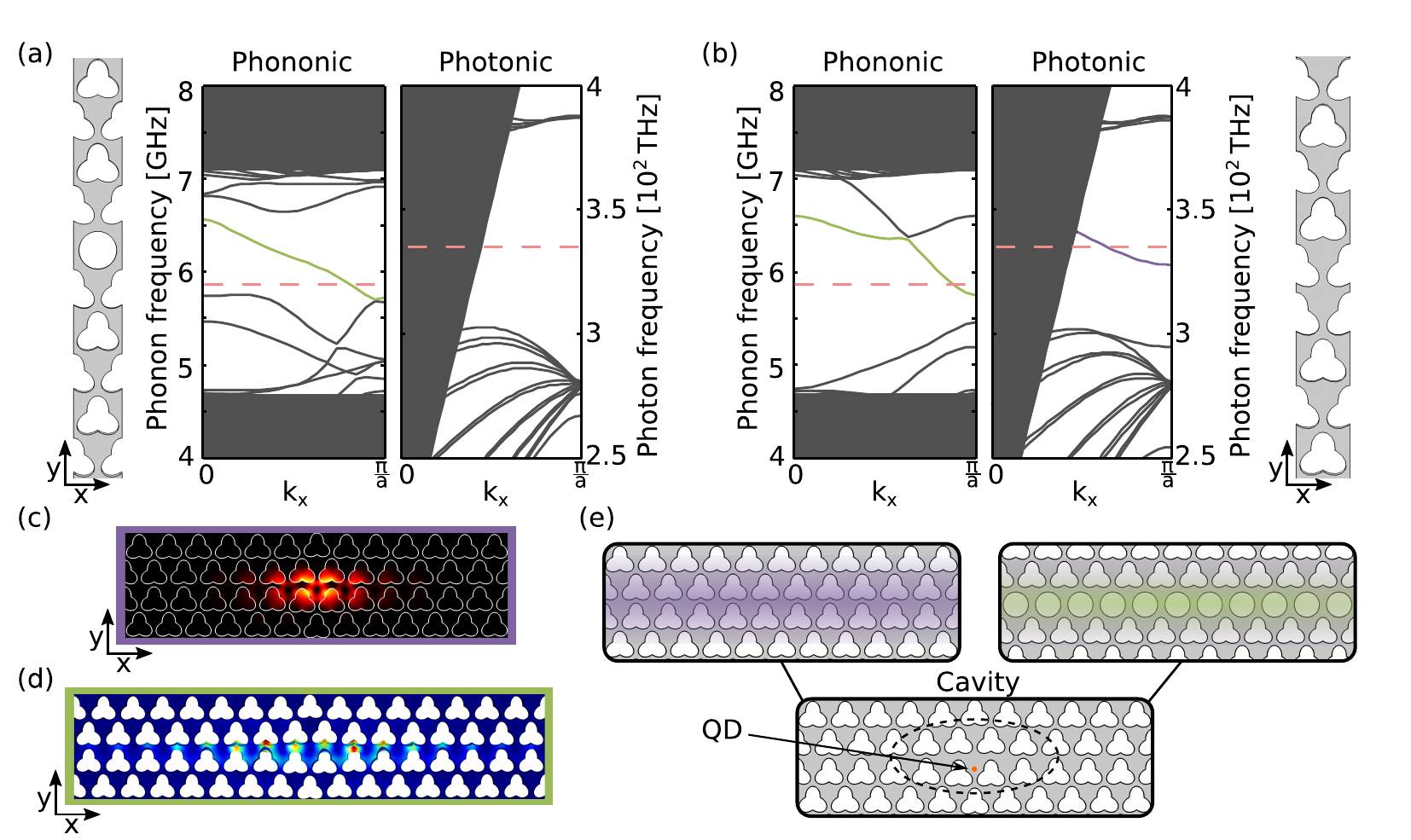}
\caption{ \label{fig:fig3} Waveguides and cavities in OM-crystal membrane structure. {\bf a)} A phonon waveguide formed by replacing a row of shamrock holes by circular holes (left). This waveguide supports a mode at the relevant phonon frequency (middle) while exhibiting a gap at the relevant photonic frequencies (right). {\bf b)} A different waveguide design, where a row of shamrock holes is removed and not replaced by anything. The gap between the two sides of the crystal is reduced to 0.58~a~$\sqrt{3}$, referred to as a W=0.58 waveguide (right). This waveguide supports both a photonic (middle) and a phononic (left) mode. The band edges of the waveguide modes (green and purple solid lines) shift up in frequency when reducing the width of the waveguide leaving the transition in a band gap for W=0.52. Combining sections of W=0.52 with a 2-period long section of W=0.58 a hetero-structure cavity is formed. {\bf c)} The modulus squared, $|E|^2$, profile of an optical resonance at $\SI{337}{\tera\hertz}$. {\bf d)} The displacement of a mechanical resonance at $\SI{5.9}{\giga\hertz}$. {\bf e)} The three opto-mechanical circuit elements needed to implement the schematic in Fig.~\ref{fig:fig1}(b). The QD is positioned in the OM-cavity coupling to the two modes shown in {\bf c)} and {\bf d)}.}
\end{figure*}
We now demonstrate some of the different types of defect modes employed for the realization of the proposal. In Fig.~\ref{fig:fig3}(a) we show a phononic waveguide formed by replacing a row of shamrock holes by circular holes. This waveguide exhibits a gap at the relevant photonic frequencies, and corresponds to the waveguide at the top of Fig.~\ref{fig:fig1}(b). It allows introducing a large $\kappa_{\textrm{e,m}}$, by bringing it close to the cavity, while leaving the optical cavity modes largely unaffected. A different waveguide design is obtained by removing a row of shamrock holes and scaling the spacing between the remaining adjacent rows of holes  to $\sqrt{3} \textrm{a W}$. The modes for a waveguide width of W=0.58 are shown in Fig.3(b). This waveguide supports both a photonic and a phononic mode with band edges close to the respective transition frequencies of the emitter. The band edges of both modes shift up in frequency when reducing the width of the waveguide, eventually, leading to a dual band gap at the frequencies relevant for the emitter for W=0.52. This makes the waveguide in Fig.3(b) well suited for the design of hetero-structure cavities, that simultaneously support modes at the desired photonic and phononic frequencies \cite{Safavi-naeini2010a, Safavi-Naeini2011}, which corresponds to the OM-cavity drawn in Fig.~\ref{fig:fig1}(b). One realization of such a cavity and the mode-profiles of the optical and the mechanical modes are shown in Fig.~\ref{fig:fig3}(c) and (d). In Fig.~\ref{fig:fig3}(e) the circuit elements needed to assemble the circuit sketched in Fig.~\ref{fig:fig1}(b) are collected. The two types of waveguides discussed suffice to realize the OM-circuit sketched in Fig.~\ref{fig:fig1}(b) and (c) by ensuring that the cavities dominant mechanical loss rate, $\kappa_{\textrm{e,m}}$, is to the phononic waveguide in Fig.~\ref{fig:fig3}(a). However, similar design ideas have been used to realize photonic waveguide defects in the same OM-crystal (not shown). Thus, this new type of OM-crystal is a versatile platform for on-chip opto-mechanics.

In this work we have demonstrated a scheme for single-phonon generation in an opto-mechanical crystal. The scheme is based on QD embedded in a membrane GaAs nanostructure whose periodic properties lead to simultaneous photonic and phononic band-gaps, allowing to control optical and acoustical interaction processes in the emitter. We have designed waveguides and cavities for both photons and phonons by introducing different types of crystal defects. Throughout the work the focus has been on designing structures that can be experimentally realized within the current scope of gallium arsenide nano-fabrication.

Finally we note that other solid-state systems, such as silicon-vacancy centers in diamond, appear to have spin-coherence times limited by single-phonon mediated relaxation processes \cite{Rogers2014b,Pingault2014}. In addition to enhancing the spin coherence of such a system it also becomes possible for the phonons to act as a coherent on-chip quantum-bus, which can couple several emitters \cite{Bennett2013} or act as a transducer from the optical to the microwave regime \cite{Davanco2012,Bochmann2013}. When exploiting the type of three-level system discussed here the resulting photon--phonon cascade can be used to implement the DLCZ-protocol (Duan, Lukin, Cirac, and Zoller)\cite{Duan2001}. Other possible applications include the creation of vibration amplification by stimulated emission of radiation \cite{Kepesidis2013}.

We would like to thank Alisa Javadi, Anders S\o rensen, Richard Warburton, Petru Tighineanu, Sahand Mahmoodian, and Tommaso Pregnolato for helpful discussions. We gratefully acknowledge financial support from the the Danish Council for Independent Research (Natural Sciences and Technology and Production Sciences) and the European Research Council (ERC Consolidator Grant - ALLQUANTUM).

\bibliographystyle{apsrev4-1}

\begin{thebibliography}{53}%
\makeatletter
\providecommand \@ifxundefined [1]{%
 \@ifx{#1\undefined}
}%
\providecommand \@ifnum [1]{%
 \ifnum #1\expandafter \@firstoftwo
 \else \expandafter \@secondoftwo
 \fi
}%
\providecommand \@ifx [1]{%
 \ifx #1\expandafter \@firstoftwo
 \else \expandafter \@secondoftwo
 \fi
}%
\providecommand \natexlab [1]{#1}%
\providecommand \enquote  [1]{``#1''}%
\providecommand \bibnamefont  [1]{#1}%
\providecommand \bibfnamefont [1]{#1}%
\providecommand \citenamefont [1]{#1}%
\providecommand \href@noop [0]{\@secondoftwo}%
\providecommand \href [0]{\begingroup \@sanitize@url \@href}%
\providecommand \@href[1]{\@@startlink{#1}\@@href}%
\providecommand \@@href[1]{\endgroup#1\@@endlink}%
\providecommand \@sanitize@url [0]{\catcode `\\12\catcode `\$12\catcode
  `\&12\catcode `\#12\catcode `\^12\catcode `\_12\catcode `\%12\relax}%
\providecommand \@@startlink[1]{}%
\providecommand \@@endlink[0]{}%
\providecommand \url  [0]{\begingroup\@sanitize@url \@url }%
\providecommand \@url [1]{\endgroup\@href {#1}{\urlprefix }}%
\providecommand \urlprefix  [0]{URL }%
\providecommand \Eprint [0]{\href }%
\providecommand \doibase [0]{http://dx.doi.org/}%
\providecommand \selectlanguage [0]{\@gobble}%
\providecommand \bibinfo  [0]{\@secondoftwo}%
\providecommand \bibfield  [0]{\@secondoftwo}%
\providecommand \translation [1]{[#1]}%
\providecommand \BibitemOpen [0]{}%
\providecommand \bibitemStop [0]{}%
\providecommand \bibitemNoStop [0]{.\EOS\space}%
\providecommand \EOS [0]{\spacefactor3000\relax}%
\providecommand \BibitemShut  [1]{\csname bibitem#1\endcsname}%
\let\auto@bib@innerbib\@empty
%</preamble>
\bibitem [{\citenamefont {Joannopoulos}\ \emph {et~al.}(2011)\citenamefont
  {Joannopoulos}, \citenamefont {Johnson}, \citenamefont {Winn},\ and\
  \citenamefont {Meade}}]{Joannopoulos2008}%
  \BibitemOpen
  \bibfield  {author} {\bibinfo {author} {\bibfnamefont {J.~D.}\ \bibnamefont
  {Joannopoulos}}, \bibinfo {author} {\bibfnamefont {S.~G.}\ \bibnamefont
  {Johnson}}, \bibinfo {author} {\bibfnamefont {J.~N.}\ \bibnamefont {Winn}}, \
  and\ \bibinfo {author} {\bibfnamefont {R.~D.}\ \bibnamefont {Meade}},\
  }\href@noop {} {\emph {\bibinfo {title} {{Photonic Crystals: Molding the Flow
  of Light}}}},\ \bibinfo {edition} {2nd}\ ed.\ (\bibinfo  {publisher}
  {Princeton University Press},\ \bibinfo {year} {2011})\BibitemShut {NoStop}%
\bibitem [{\citenamefont {Lodahl}\ \emph {et~al.}(2015)\citenamefont {Lodahl},
  \citenamefont {Mahmoodian},\ and\ \citenamefont {Stobbe}}]{Lodahl2013}%
  \BibitemOpen
  \bibfield  {author} {\bibinfo {author} {\bibfnamefont {P.}~\bibnamefont
  {Lodahl}}, \bibinfo {author} {\bibfnamefont {S.}~\bibnamefont {Mahmoodian}},
  \ and\ \bibinfo {author} {\bibfnamefont {S.}~\bibnamefont {Stobbe}},\ }\href
  {\doibase 10.1103/PhysRevB.49.14352} {\bibfield  {journal} {\bibinfo
  {journal} {Reviews of Modern Physics}\ }\textbf {\bibinfo {volume} {87}},\
  \bibinfo {pages} {347} (\bibinfo {year} {2015})}\BibitemShut {NoStop}%
\bibitem [{\citenamefont {Gorishnyy}\ \emph {et~al.}(2005)\citenamefont
  {Gorishnyy}, \citenamefont {Maldovan}, \citenamefont {Ullal},\ and\
  \citenamefont {Thomas}}]{Gorishnyy2005a}%
  \BibitemOpen
  \bibfield  {author} {\bibinfo {author} {\bibfnamefont {T.}~\bibnamefont
  {Gorishnyy}}, \bibinfo {author} {\bibfnamefont {M.}~\bibnamefont {Maldovan}},
  \bibinfo {author} {\bibfnamefont {C.}~\bibnamefont {Ullal}}, \ and\ \bibinfo
  {author} {\bibfnamefont {E.}~\bibnamefont {Thomas}},\ }\href@noop {}
  {\bibfield  {journal} {\bibinfo  {journal} {Physics World}\ }\textbf
  {\bibinfo {volume} {18}},\ \bibinfo {pages} {24} (\bibinfo {year}
  {2005})}\BibitemShut {NoStop}%
\bibitem [{\citenamefont {Maldovan}\ and\ \citenamefont
  {Thomas}(2006)}]{Maldovan2006}%
  \BibitemOpen
  \bibfield  {author} {\bibinfo {author} {\bibfnamefont {M.}~\bibnamefont
  {Maldovan}}\ and\ \bibinfo {author} {\bibfnamefont {E.~L.}\ \bibnamefont
  {Thomas}},\ }\href {\doibase 10.1063/1.2216885} {\bibfield  {journal}
  {\bibinfo  {journal} {Applied Physics Letters}\ }\textbf {\bibinfo {volume}
  {88}},\ \bibinfo {pages} {251907} (\bibinfo {year} {2006})}\BibitemShut
  {NoStop}%
\bibitem [{\citenamefont {Eichenfield}\ \emph {et~al.}(2009)\citenamefont
  {Eichenfield}, \citenamefont {Chan}, \citenamefont {Camacho}, \citenamefont
  {Vahala},\ and\ \citenamefont {Painter}}]{Eichenfield2009}%
  \BibitemOpen
  \bibfield  {author} {\bibinfo {author} {\bibfnamefont {M.}~\bibnamefont
  {Eichenfield}}, \bibinfo {author} {\bibfnamefont {J.}~\bibnamefont {Chan}},
  \bibinfo {author} {\bibfnamefont {R.~M.}\ \bibnamefont {Camacho}}, \bibinfo
  {author} {\bibfnamefont {K.~J.}\ \bibnamefont {Vahala}}, \ and\ \bibinfo
  {author} {\bibfnamefont {O.}~\bibnamefont {Painter}},\ }\href {\doibase
  10.1038/nature08524} {\bibfield  {journal} {\bibinfo  {journal} {Nature}\
  }\textbf {\bibinfo {volume} {462}},\ \bibinfo {pages} {78} (\bibinfo {year}
  {2009})}\BibitemShut {NoStop}%
\bibitem [{\citenamefont {Safavi-Naeini}\ and\ \citenamefont
  {Painter}(2010)}]{Safavi-naeini2010a}%
  \BibitemOpen
  \bibfield  {author} {\bibinfo {author} {\bibfnamefont {A.~H.}\ \bibnamefont
  {Safavi-Naeini}}\ and\ \bibinfo {author} {\bibfnamefont {O.}~\bibnamefont
  {Painter}},\ }\href {\doibase 10.1364/OE.18.014926} {\bibfield  {journal}
  {\bibinfo  {journal} {Optics Express}\ }\textbf {\bibinfo {volume} {18}},\
  \bibinfo {pages} {14926} (\bibinfo {year} {2010})}\BibitemShut {NoStop}%
\bibitem [{\citenamefont {Mohammadi}\ \emph {et~al.}(2010)\citenamefont
  {Mohammadi}, \citenamefont {Eftekhar}, \citenamefont {Khelif},\ and\
  \citenamefont {Adibi}}]{Mohammadi2010}%
  \BibitemOpen
  \bibfield  {author} {\bibinfo {author} {\bibfnamefont {S.}~\bibnamefont
  {Mohammadi}}, \bibinfo {author} {\bibfnamefont {A.~A.}\ \bibnamefont
  {Eftekhar}}, \bibinfo {author} {\bibfnamefont {A.}~\bibnamefont {Khelif}}, \
  and\ \bibinfo {author} {\bibfnamefont {A.}~\bibnamefont {Adibi}},\ }\href
  {\doibase 10.1364/OE.18.009164} {\bibfield  {journal} {\bibinfo  {journal}
  {Optics Express}\ }\textbf {\bibinfo {volume} {18}},\ \bibinfo {pages} {9164}
  (\bibinfo {year} {2010})}\BibitemShut {NoStop}%
\bibitem [{\citenamefont {Safavi-Naeini}\ and\ \citenamefont
  {Painter}(2011)}]{Safavi-Naeini2011}%
  \BibitemOpen
  \bibfield  {author} {\bibinfo {author} {\bibfnamefont {A.~H.}\ \bibnamefont
  {Safavi-Naeini}}\ and\ \bibinfo {author} {\bibfnamefont {O.}~\bibnamefont
  {Painter}},\ }\href {\doibase 10.1088/1367-2630/13/1/013017} {\bibfield
  {journal} {\bibinfo  {journal} {New Journal of Physics}\ }\textbf {\bibinfo
  {volume} {13}},\ \bibinfo {pages} {013017} (\bibinfo {year}
  {2011})}\BibitemShut {NoStop}%
\bibitem [{\citenamefont {Galland}\ \emph {et~al.}(2014)\citenamefont
  {Galland}, \citenamefont {Sangouard}, \citenamefont {Piro}, \citenamefont
  {Gisin},\ and\ \citenamefont {Kippenberg}}]{Galland2014}%
  \BibitemOpen
  \bibfield  {author} {\bibinfo {author} {\bibfnamefont {C.}~\bibnamefont
  {Galland}}, \bibinfo {author} {\bibfnamefont {N.}~\bibnamefont {Sangouard}},
  \bibinfo {author} {\bibfnamefont {N.}~\bibnamefont {Piro}}, \bibinfo {author}
  {\bibfnamefont {N.}~\bibnamefont {Gisin}}, \ and\ \bibinfo {author}
  {\bibfnamefont {T.~J.}\ \bibnamefont {Kippenberg}},\ }\href {\doibase
  10.1103/PhysRevLett.112.143602} {\bibfield  {journal} {\bibinfo  {journal}
  {Physical Review Letters}\ }\textbf {\bibinfo {volume} {112}},\ \bibinfo
  {pages} {143602} (\bibinfo {year} {2014})}\BibitemShut {NoStop}%
\bibitem [{\citenamefont {Schmidt}\ \emph {et~al.}(2015)\citenamefont
  {Schmidt}, \citenamefont {Peano},\ and\ \citenamefont
  {Marquardt}}]{Schmidt2015}%
  \BibitemOpen
  \bibfield  {author} {\bibinfo {author} {\bibfnamefont {M.}~\bibnamefont
  {Schmidt}}, \bibinfo {author} {\bibfnamefont {V.}~\bibnamefont {Peano}}, \
  and\ \bibinfo {author} {\bibfnamefont {F.}~\bibnamefont {Marquardt}},\ }\href
  {\doibase 10.1088/1367-2630/17/2/023025} {\bibfield  {journal} {\bibinfo
  {journal} {New Journal of Physics}\ }\textbf {\bibinfo {volume} {17}},\
  \bibinfo {pages} {023025} (\bibinfo {year} {2015})}\BibitemShut {NoStop}%
\bibitem [{\citenamefont {Chan}\ \emph {et~al.}(2011)\citenamefont {Chan},
  \citenamefont {Alegre}, \citenamefont {Safavi-Naeini}, \citenamefont {Hill},
  \citenamefont {Krause}, \citenamefont {Groeblacher}, \citenamefont
  {Aspelmeyer},\ and\ \citenamefont {Painter}}]{Chan2011}%
  \BibitemOpen
  \bibfield  {author} {\bibinfo {author} {\bibfnamefont {J.}~\bibnamefont
  {Chan}}, \bibinfo {author} {\bibfnamefont {T.~P.~M.}\ \bibnamefont {Alegre}},
  \bibinfo {author} {\bibfnamefont {A.~H.}\ \bibnamefont {Safavi-Naeini}},
  \bibinfo {author} {\bibfnamefont {J.~T.}\ \bibnamefont {Hill}}, \bibinfo
  {author} {\bibfnamefont {A.}~\bibnamefont {Krause}}, \bibinfo {author}
  {\bibfnamefont {S.}~\bibnamefont {Groeblacher}}, \bibinfo {author}
  {\bibfnamefont {M.}~\bibnamefont {Aspelmeyer}}, \ and\ \bibinfo {author}
  {\bibfnamefont {O.}~\bibnamefont {Painter}},\ }\href {\doibase
  10.1038/nature10461} {\bibfield  {journal} {\bibinfo  {journal} {Nature}\
  }\textbf {\bibinfo {volume} {478}},\ \bibinfo {pages} {89} (\bibinfo {year}
  {2011})}\BibitemShut {NoStop}%
\bibitem [{\citenamefont {Cohen}\ \emph {et~al.}(2015)\citenamefont {Cohen},
  \citenamefont {Meenehan}, \citenamefont {MacCabe}, \citenamefont
  {Gr{\"{o}}blacher}, \citenamefont {Safavi-Naeini}, \citenamefont {Marsili},
  \citenamefont {Shaw},\ and\ \citenamefont {Painter}}]{Cohen2015}%
  \BibitemOpen
  \bibfield  {author} {\bibinfo {author} {\bibfnamefont {J.~D.}\ \bibnamefont
  {Cohen}}, \bibinfo {author} {\bibfnamefont {S.~M.}\ \bibnamefont {Meenehan}},
  \bibinfo {author} {\bibfnamefont {G.~S.}\ \bibnamefont {MacCabe}}, \bibinfo
  {author} {\bibfnamefont {S.}~\bibnamefont {Gr{\"{o}}blacher}}, \bibinfo
  {author} {\bibfnamefont {A.~H.}\ \bibnamefont {Safavi-Naeini}}, \bibinfo
  {author} {\bibfnamefont {F.}~\bibnamefont {Marsili}}, \bibinfo {author}
  {\bibfnamefont {M.~D.}\ \bibnamefont {Shaw}}, \ and\ \bibinfo {author}
  {\bibfnamefont {O.}~\bibnamefont {Painter}},\ }\href {\doibase
  10.1038/nature14349} {\bibfield  {journal} {\bibinfo  {journal} {Nature}\
  }\textbf {\bibinfo {volume} {520}},\ \bibinfo {pages} {522} (\bibinfo {year}
  {2015})}\BibitemShut {NoStop}%
\bibitem [{\citenamefont {Riedinger}\ \emph {et~al.}(2016)\citenamefont
  {Riedinger}, \citenamefont {Hong}, \citenamefont {Norte}, \citenamefont
  {Slater}, \citenamefont {Shang}, \citenamefont {Krause}, \citenamefont
  {Anant}, \citenamefont {Aspelmeyer},\ and\ \citenamefont
  {Gr{\"{o}}blacher}}]{Riedinger2015}%
  \BibitemOpen
  \bibfield  {author} {\bibinfo {author} {\bibfnamefont {R.}~\bibnamefont
  {Riedinger}}, \bibinfo {author} {\bibfnamefont {S.}~\bibnamefont {Hong}},
  \bibinfo {author} {\bibfnamefont {R.~A.}\ \bibnamefont {Norte}}, \bibinfo
  {author} {\bibfnamefont {J.~A.}\ \bibnamefont {Slater}}, \bibinfo {author}
  {\bibfnamefont {J.}~\bibnamefont {Shang}}, \bibinfo {author} {\bibfnamefont
  {A.~G.}\ \bibnamefont {Krause}}, \bibinfo {author} {\bibfnamefont
  {V.}~\bibnamefont {Anant}}, \bibinfo {author} {\bibfnamefont
  {M.}~\bibnamefont {Aspelmeyer}}, \ and\ \bibinfo {author} {\bibfnamefont
  {S.}~\bibnamefont {Gr{\"{o}}blacher}},\ }\href {\doibase 10.1038/nature16536}
  {\bibfield  {journal} {\bibinfo  {journal} {Nature}\ } (\bibinfo {year}
  {2016}),\ 10.1038/nature16536}\BibitemShut {NoStop}%
\bibitem [{\citenamefont {Rabl}\ \emph {et~al.}(2009)\citenamefont {Rabl},
  \citenamefont {Cappellaro}, \citenamefont {Gurudev~Dutt}, \citenamefont
  {Jiang}, \citenamefont {Maze},\ and\ \citenamefont {Lukin}}]{Rabl2009}%
  \BibitemOpen
  \bibfield  {author} {\bibinfo {author} {\bibfnamefont {P.}~\bibnamefont
  {Rabl}}, \bibinfo {author} {\bibfnamefont {P.}~\bibnamefont {Cappellaro}},
  \bibinfo {author} {\bibfnamefont {M.~V.}\ \bibnamefont {Gurudev~Dutt}},
  \bibinfo {author} {\bibfnamefont {L.}~\bibnamefont {Jiang}}, \bibinfo
  {author} {\bibfnamefont {J.~R.}\ \bibnamefont {Maze}}, \ and\ \bibinfo
  {author} {\bibfnamefont {M.~D.}\ \bibnamefont {Lukin}},\ }\href {\doibase
  10.1103/PhysRevB.79.041302} {\bibfield  {journal} {\bibinfo  {journal}
  {Physical Review B}\ }\textbf {\bibinfo {volume} {79}},\ \bibinfo {pages}
  {041302} (\bibinfo {year} {2009})}\BibitemShut {NoStop}%
\bibitem [{\citenamefont {Arcizet}\ \emph {et~al.}(2011)\citenamefont
  {Arcizet}, \citenamefont {Jacques}, \citenamefont {Siria}, \citenamefont
  {Poncharal}, \citenamefont {Vincent},\ and\ \citenamefont
  {Seidelin}}]{Arcizet2011}%
  \BibitemOpen
  \bibfield  {author} {\bibinfo {author} {\bibfnamefont {O.}~\bibnamefont
  {Arcizet}}, \bibinfo {author} {\bibfnamefont {V.}~\bibnamefont {Jacques}},
  \bibinfo {author} {\bibfnamefont {A.}~\bibnamefont {Siria}}, \bibinfo
  {author} {\bibfnamefont {P.}~\bibnamefont {Poncharal}}, \bibinfo {author}
  {\bibfnamefont {P.}~\bibnamefont {Vincent}}, \ and\ \bibinfo {author}
  {\bibfnamefont {S.}~\bibnamefont {Seidelin}},\ }\href {\doibase
  10.1038/nphys2070} {\bibfield  {journal} {\bibinfo  {journal} {Nature
  Physics}\ }\textbf {\bibinfo {volume} {7}},\ \bibinfo {pages} {879} (\bibinfo
  {year} {2011})}\BibitemShut {NoStop}%
\bibitem [{\citenamefont {Wilson-Rae}\ \emph {et~al.}(2004)\citenamefont
  {Wilson-Rae}, \citenamefont {Zoller},\ and\ \citenamefont
  {Imamo\u{g}lu}}]{Wilson-Rae2004}%
  \BibitemOpen
  \bibfield  {author} {\bibinfo {author} {\bibfnamefont {I.}~\bibnamefont
  {Wilson-Rae}}, \bibinfo {author} {\bibfnamefont {P.}~\bibnamefont {Zoller}},
  \ and\ \bibinfo {author} {\bibfnamefont {A.}~\bibnamefont {Imamo\u{g}lu}},\
  }\href {\doibase 10.1103/PhysRevLett.92.075507} {\bibfield  {journal}
  {\bibinfo  {journal} {Physical Review Letters}\ }\textbf {\bibinfo {volume}
  {92}},\ \bibinfo {pages} {075507} (\bibinfo {year} {2004})}\BibitemShut
  {NoStop}%
\bibitem [{\citenamefont {Bennett}\ \emph {et~al.}(2013)\citenamefont
  {Bennett}, \citenamefont {Yao}, \citenamefont {Otterbach}, \citenamefont
  {Zoller}, \citenamefont {Rabl},\ and\ \citenamefont {Lukin}}]{Bennett2013}%
  \BibitemOpen
  \bibfield  {author} {\bibinfo {author} {\bibfnamefont {S.~D.}\ \bibnamefont
  {Bennett}}, \bibinfo {author} {\bibfnamefont {N.~Y.}\ \bibnamefont {Yao}},
  \bibinfo {author} {\bibfnamefont {J.}~\bibnamefont {Otterbach}}, \bibinfo
  {author} {\bibfnamefont {P.}~\bibnamefont {Zoller}}, \bibinfo {author}
  {\bibfnamefont {P.}~\bibnamefont {Rabl}}, \ and\ \bibinfo {author}
  {\bibfnamefont {M.~D.}\ \bibnamefont {Lukin}},\ }\href {\doibase
  10.1103/PhysRevLett.110.156402} {\bibfield  {journal} {\bibinfo  {journal}
  {Physical Review Letters}\ }\textbf {\bibinfo {volume} {110}},\ \bibinfo
  {pages} {156402} (\bibinfo {year} {2013})}\BibitemShut {NoStop}%
\bibitem [{\citenamefont {Ovartchaiyapong}\ \emph {et~al.}(2014)\citenamefont
  {Ovartchaiyapong}, \citenamefont {Lee}, \citenamefont {Myers},\ and\
  \citenamefont {Jayich}}]{Ovartchaiyapong2014}%
  \BibitemOpen
  \bibfield  {author} {\bibinfo {author} {\bibfnamefont {P.}~\bibnamefont
  {Ovartchaiyapong}}, \bibinfo {author} {\bibfnamefont {K.~W.}\ \bibnamefont
  {Lee}}, \bibinfo {author} {\bibfnamefont {B.~A.}\ \bibnamefont {Myers}}, \
  and\ \bibinfo {author} {\bibfnamefont {A.~C.~B.}\ \bibnamefont {Jayich}},\
  }\href {\doibase 10.1038/ncomms5429} {\bibfield  {journal} {\bibinfo
  {journal} {Nature Communications}\ }\textbf {\bibinfo {volume} {5}},\
  \bibinfo {pages} {4429} (\bibinfo {year} {2014})}\BibitemShut {NoStop}%
\bibitem [{\citenamefont {Barfuss}\ \emph {et~al.}(2015)\citenamefont
  {Barfuss}, \citenamefont {Teissier}, \citenamefont {Neu}, \citenamefont
  {Nunnenkamp},\ and\ \citenamefont {Maletinsky}}]{Barfuss2015}%
  \BibitemOpen
  \bibfield  {author} {\bibinfo {author} {\bibfnamefont {A.}~\bibnamefont
  {Barfuss}}, \bibinfo {author} {\bibfnamefont {J.}~\bibnamefont {Teissier}},
  \bibinfo {author} {\bibfnamefont {E.}~\bibnamefont {Neu}}, \bibinfo {author}
  {\bibfnamefont {A.}~\bibnamefont {Nunnenkamp}}, \ and\ \bibinfo {author}
  {\bibfnamefont {P.}~\bibnamefont {Maletinsky}},\ }\href {\doibase
  10.1038/nphys3411} {\bibfield  {journal} {\bibinfo  {journal} {Nature
  Physics}\ }\textbf {\bibinfo {volume} {11}},\ \bibinfo {pages} {820}
  (\bibinfo {year} {2015})}\BibitemShut {NoStop}%
\bibitem [{\citenamefont {MacQuarrie}\ \emph {et~al.}(2013)\citenamefont
  {MacQuarrie}, \citenamefont {Gosavi}, \citenamefont {Jungwirth},
  \citenamefont {Bhave},\ and\ \citenamefont {Fuchs}}]{MacQuarrie2013}%
  \BibitemOpen
  \bibfield  {author} {\bibinfo {author} {\bibfnamefont {E.~R.}\ \bibnamefont
  {MacQuarrie}}, \bibinfo {author} {\bibfnamefont {T.~A.}\ \bibnamefont
  {Gosavi}}, \bibinfo {author} {\bibfnamefont {N.~R.}\ \bibnamefont
  {Jungwirth}}, \bibinfo {author} {\bibfnamefont {S.~A.}\ \bibnamefont
  {Bhave}}, \ and\ \bibinfo {author} {\bibfnamefont {G.~D.}\ \bibnamefont
  {Fuchs}},\ }\href {\doibase 10.1103/PhysRevLett.111.227602} {\bibfield
  {journal} {\bibinfo  {journal} {Physical Review Letters}\ }\textbf {\bibinfo
  {volume} {111}},\ \bibinfo {pages} {227602} (\bibinfo {year}
  {2013})}\BibitemShut {NoStop}%
\bibitem [{\citenamefont {Yeo}\ \emph {et~al.}(2013)\citenamefont {Yeo},
  \citenamefont {de~Assis}, \citenamefont {Gloppe}, \citenamefont
  {Dupont-Ferrier}, \citenamefont {Verlot}, \citenamefont {Malik},
  \citenamefont {Dupuy}, \citenamefont {Claudon}, \citenamefont {G{\'{e}}rard},
  \citenamefont {Auff{\`{e}}ves}, \citenamefont {Nogues}, \citenamefont
  {Seidelin}, \citenamefont {Poizat}, \citenamefont {Arcizet},\ and\
  \citenamefont {Richard}}]{Yeo2013}%
  \BibitemOpen
  \bibfield  {author} {\bibinfo {author} {\bibfnamefont {I.}~\bibnamefont
  {Yeo}}, \bibinfo {author} {\bibfnamefont {P.-L.}\ \bibnamefont {de~Assis}},
  \bibinfo {author} {\bibfnamefont {A.}~\bibnamefont {Gloppe}}, \bibinfo
  {author} {\bibfnamefont {E.}~\bibnamefont {Dupont-Ferrier}}, \bibinfo
  {author} {\bibfnamefont {P.}~\bibnamefont {Verlot}}, \bibinfo {author}
  {\bibfnamefont {N.~S.}\ \bibnamefont {Malik}}, \bibinfo {author}
  {\bibfnamefont {E.}~\bibnamefont {Dupuy}}, \bibinfo {author} {\bibfnamefont
  {J.}~\bibnamefont {Claudon}}, \bibinfo {author} {\bibfnamefont {J.-M.}\
  \bibnamefont {G{\'{e}}rard}}, \bibinfo {author} {\bibfnamefont
  {A.}~\bibnamefont {Auff{\`{e}}ves}}, \bibinfo {author} {\bibfnamefont
  {G.}~\bibnamefont {Nogues}}, \bibinfo {author} {\bibfnamefont
  {S.}~\bibnamefont {Seidelin}}, \bibinfo {author} {\bibfnamefont {J.-P.}\
  \bibnamefont {Poizat}}, \bibinfo {author} {\bibfnamefont {O.}~\bibnamefont
  {Arcizet}}, \ and\ \bibinfo {author} {\bibfnamefont {M.}~\bibnamefont
  {Richard}},\ }\href {\doibase 10.1038/nnano.2013.274} {\bibfield  {journal}
  {\bibinfo  {journal} {Nature Nanotechnology}\ }\textbf {\bibinfo {volume}
  {9}},\ \bibinfo {pages} {106} (\bibinfo {year} {2013})}\BibitemShut {NoStop}%
\bibitem [{\citenamefont {Montinaro}\ \emph {et~al.}(2014)\citenamefont
  {Montinaro}, \citenamefont {W{\"{u}}st}, \citenamefont {Munsch},
  \citenamefont {Fontana}, \citenamefont {Russo-Averchi}, \citenamefont
  {Heiss}, \citenamefont {{Fontcuberta I Morral}}, \citenamefont {Warburton},\
  and\ \citenamefont {Poggio}}]{Montinaro2014}%
  \BibitemOpen
  \bibfield  {author} {\bibinfo {author} {\bibfnamefont {M.}~\bibnamefont
  {Montinaro}}, \bibinfo {author} {\bibfnamefont {G.}~\bibnamefont
  {W{\"{u}}st}}, \bibinfo {author} {\bibfnamefont {M.}~\bibnamefont {Munsch}},
  \bibinfo {author} {\bibfnamefont {Y.}~\bibnamefont {Fontana}}, \bibinfo
  {author} {\bibfnamefont {E.}~\bibnamefont {Russo-Averchi}}, \bibinfo {author}
  {\bibfnamefont {M.}~\bibnamefont {Heiss}}, \bibinfo {author} {\bibfnamefont
  {A.}~\bibnamefont {{Fontcuberta I Morral}}}, \bibinfo {author} {\bibfnamefont
  {R.~J.}\ \bibnamefont {Warburton}}, \ and\ \bibinfo {author} {\bibfnamefont
  {M.}~\bibnamefont {Poggio}},\ }\href {\doibase 10.1021/nl501413t} {\bibfield
  {journal} {\bibinfo  {journal} {Nano Letters}\ }\textbf {\bibinfo {volume}
  {14}},\ \bibinfo {pages} {4454} (\bibinfo {year} {2014})}\BibitemShut
  {NoStop}%
\bibitem [{\citenamefont {Colless}\ \emph {et~al.}(2014)\citenamefont
  {Colless}, \citenamefont {Croot}, \citenamefont {Stace}, \citenamefont
  {Doherty}, \citenamefont {Barrett}, \citenamefont {Lu}, \citenamefont
  {Gossard},\ and\ \citenamefont {Reilly}}]{Colless2014}%
  \BibitemOpen
  \bibfield  {author} {\bibinfo {author} {\bibfnamefont {J.~I.}\ \bibnamefont
  {Colless}}, \bibinfo {author} {\bibfnamefont {X.~G.}\ \bibnamefont {Croot}},
  \bibinfo {author} {\bibfnamefont {T.~M.}\ \bibnamefont {Stace}}, \bibinfo
  {author} {\bibfnamefont {A.~C.}\ \bibnamefont {Doherty}}, \bibinfo {author}
  {\bibfnamefont {S.~D.}\ \bibnamefont {Barrett}}, \bibinfo {author}
  {\bibfnamefont {H.}~\bibnamefont {Lu}}, \bibinfo {author} {\bibfnamefont
  {A.~C.}\ \bibnamefont {Gossard}}, \ and\ \bibinfo {author} {\bibfnamefont
  {D.~J.}\ \bibnamefont {Reilly}},\ }\href {\doibase 10.1038/ncomms4716}
  {\bibfield  {journal} {\bibinfo  {journal} {Nature Communications}\ }\textbf
  {\bibinfo {volume} {5}},\ \bibinfo {pages} {3716} (\bibinfo {year}
  {2014})}\BibitemShut {NoStop}%
\bibitem [{\citenamefont {Jahnke}\ \emph {et~al.}(2015)\citenamefont {Jahnke},
  \citenamefont {Sipahigil}, \citenamefont {Binder}, \citenamefont {Doherty},
  \citenamefont {Metsch}, \citenamefont {Rogers}, \citenamefont {Manson},
  \citenamefont {Lukin},\ and\ \citenamefont {Jelezko}}]{Jahnke2015}%
  \BibitemOpen
  \bibfield  {author} {\bibinfo {author} {\bibfnamefont {K.~D.}\ \bibnamefont
  {Jahnke}}, \bibinfo {author} {\bibfnamefont {A.}~\bibnamefont {Sipahigil}},
  \bibinfo {author} {\bibfnamefont {J.~M.}\ \bibnamefont {Binder}}, \bibinfo
  {author} {\bibfnamefont {M.~W.}\ \bibnamefont {Doherty}}, \bibinfo {author}
  {\bibfnamefont {M.}~\bibnamefont {Metsch}}, \bibinfo {author} {\bibfnamefont
  {L.~J.}\ \bibnamefont {Rogers}}, \bibinfo {author} {\bibfnamefont {N.~B.}\
  \bibnamefont {Manson}}, \bibinfo {author} {\bibfnamefont {M.~D.}\
  \bibnamefont {Lukin}}, \ and\ \bibinfo {author} {\bibfnamefont
  {F.}~\bibnamefont {Jelezko}},\ }\href {\doibase
  10.1088/1367-2630/17/4/043011} {\bibfield  {journal} {\bibinfo  {journal}
  {New Journal of Physics}\ }\textbf {\bibinfo {volume} {17}},\ \bibinfo
  {pages} {043011} (\bibinfo {year} {2015})}\BibitemShut {NoStop}%
\bibitem [{\citenamefont {Flayac}\ and\ \citenamefont
  {Savona}(2014)}]{Flayac2014}%
  \BibitemOpen
  \bibfield  {author} {\bibinfo {author} {\bibfnamefont {H.}~\bibnamefont
  {Flayac}}\ and\ \bibinfo {author} {\bibfnamefont {V.}~\bibnamefont
  {Savona}},\ }\href {\doibase 10.1103/PhysRevLett.113.143603} {\bibfield
  {journal} {\bibinfo  {journal} {Physical Review Letters}\ }\textbf {\bibinfo
  {volume} {113}},\ \bibinfo {pages} {143603} (\bibinfo {year}
  {2014})}\BibitemShut {NoStop}%
\bibitem [{\citenamefont {Aspelmeyer}\ \emph {et~al.}(2014)\citenamefont
  {Aspelmeyer}, \citenamefont {Kippenberg},\ and\ \citenamefont
  {Marquardt}}]{Aspelmeyer2014}%
  \BibitemOpen
  \bibfield  {author} {\bibinfo {author} {\bibfnamefont {M.}~\bibnamefont
  {Aspelmeyer}}, \bibinfo {author} {\bibfnamefont {T.~J.}\ \bibnamefont
  {Kippenberg}}, \ and\ \bibinfo {author} {\bibfnamefont {F.}~\bibnamefont
  {Marquardt}},\ }\href {\doibase 10.1103/RevModPhys.86.1391} {\bibfield
  {journal} {\bibinfo  {journal} {Reviews of Modern Physics}\ }\textbf
  {\bibinfo {volume} {86}},\ \bibinfo {pages} {1391} (\bibinfo {year}
  {2014})}\BibitemShut {NoStop}%
\bibitem [{\citenamefont {Meenehan}\ \emph {et~al.}(2014)\citenamefont
  {Meenehan}, \citenamefont {Cohen}, \citenamefont {Gr{\"{o}}blacher},
  \citenamefont {Hill}, \citenamefont {Safavi-Naeini}, \citenamefont
  {Aspelmeyer},\ and\ \citenamefont {Painter}}]{Meenehan2014}%
  \BibitemOpen
  \bibfield  {author} {\bibinfo {author} {\bibfnamefont {S.~M.}\ \bibnamefont
  {Meenehan}}, \bibinfo {author} {\bibfnamefont {J.~D.}\ \bibnamefont {Cohen}},
  \bibinfo {author} {\bibfnamefont {S.}~\bibnamefont {Gr{\"{o}}blacher}},
  \bibinfo {author} {\bibfnamefont {J.~T.}\ \bibnamefont {Hill}}, \bibinfo
  {author} {\bibfnamefont {A.~H.}\ \bibnamefont {Safavi-Naeini}}, \bibinfo
  {author} {\bibfnamefont {M.}~\bibnamefont {Aspelmeyer}}, \ and\ \bibinfo
  {author} {\bibfnamefont {O.}~\bibnamefont {Painter}},\ }\href {\doibase
  10.1103/PhysRevA.90.011803} {\bibfield  {journal} {\bibinfo  {journal}
  {Physical Review A}\ }\textbf {\bibinfo {volume} {90}},\ \bibinfo {pages}
  {011803(R)} (\bibinfo {year} {2014})}\BibitemShut {NoStop}%
\bibitem [{Note1()}]{Note1}%
  \BibitemOpen
  \bibinfo {note} {\label {footnote1}We refer to Ref.~\cite {Sollner2015} for a
  discussion about what constitutes narrow bandwidth for the incident photon
  and the role of dephasing in the coherent scattering regime.}\BibitemShut
  {Stop}%
\bibitem [{\citenamefont {Press}\ \emph {et~al.}(2008)\citenamefont {Press},
  \citenamefont {Ladd}, \citenamefont {Zhang},\ and\ \citenamefont
  {Yamamoto}}]{Press2008}%
  \BibitemOpen
  \bibfield  {author} {\bibinfo {author} {\bibfnamefont {D.}~\bibnamefont
  {Press}}, \bibinfo {author} {\bibfnamefont {T.~D.}\ \bibnamefont {Ladd}},
  \bibinfo {author} {\bibfnamefont {B.}~\bibnamefont {Zhang}}, \ and\ \bibinfo
  {author} {\bibfnamefont {Y.}~\bibnamefont {Yamamoto}},\ }\href {\doibase
  10.1038/nature07530} {\bibfield  {journal} {\bibinfo  {journal} {Nature}\
  }\textbf {\bibinfo {volume} {456}},\ \bibinfo {pages} {218} (\bibinfo {year}
  {2008})}\BibitemShut {NoStop}%
\bibitem [{\citenamefont {Kroutvar}\ \emph {et~al.}(2004)\citenamefont
  {Kroutvar}, \citenamefont {Ducommun}, \citenamefont {Heiss}, \citenamefont
  {Bichler}, \citenamefont {Schuh}, \citenamefont {Abstreiter},\ and\
  \citenamefont {Finley}}]{Kroutvar2004}%
  \BibitemOpen
  \bibfield  {author} {\bibinfo {author} {\bibfnamefont {M.}~\bibnamefont
  {Kroutvar}}, \bibinfo {author} {\bibfnamefont {Y.}~\bibnamefont {Ducommun}},
  \bibinfo {author} {\bibfnamefont {D.}~\bibnamefont {Heiss}}, \bibinfo
  {author} {\bibfnamefont {M.}~\bibnamefont {Bichler}}, \bibinfo {author}
  {\bibfnamefont {D.}~\bibnamefont {Schuh}}, \bibinfo {author} {\bibfnamefont
  {G.}~\bibnamefont {Abstreiter}}, \ and\ \bibinfo {author} {\bibfnamefont
  {J.~J.}\ \bibnamefont {Finley}},\ }\href {\doibase 10.1038/nature02986.}
  {\bibfield  {journal} {\bibinfo  {journal} {Nature}\ }\textbf {\bibinfo
  {volume} {432}},\ \bibinfo {pages} {81} (\bibinfo {year} {2004})}\BibitemShut
  {NoStop}%
\bibitem [{\citenamefont {Dreiser}\ \emph {et~al.}(2008)\citenamefont
  {Dreiser}, \citenamefont {Atat{\"{u}}re}, \citenamefont {Galland},
  \citenamefont {M{\"{u}}ller}, \citenamefont {Badolato},\ and\ \citenamefont
  {Imamo\u{g}lu}}]{Dreiser2008}%
  \BibitemOpen
  \bibfield  {author} {\bibinfo {author} {\bibfnamefont {J.}~\bibnamefont
  {Dreiser}}, \bibinfo {author} {\bibfnamefont {M.}~\bibnamefont
  {Atat{\"{u}}re}}, \bibinfo {author} {\bibfnamefont {C.}~\bibnamefont
  {Galland}}, \bibinfo {author} {\bibfnamefont {T.}~\bibnamefont
  {M{\"{u}}ller}}, \bibinfo {author} {\bibfnamefont {A.}~\bibnamefont
  {Badolato}}, \ and\ \bibinfo {author} {\bibfnamefont {A.}~\bibnamefont
  {Imamo\u{g}lu}},\ }\href {\doibase 10.1103/PhysRevB.77.075317} {\bibfield
  {journal} {\bibinfo  {journal} {Physical Review B}\ }\textbf {\bibinfo
  {volume} {77}},\ \bibinfo {pages} {075317} (\bibinfo {year}
  {2008})}\BibitemShut {NoStop}%
\bibitem [{\citenamefont {Khaetskii}\ and\ \citenamefont
  {Nazarov}(2001)}]{Khaetskii2001}%
  \BibitemOpen
  \bibfield  {author} {\bibinfo {author} {\bibfnamefont {A.~V.}\ \bibnamefont
  {Khaetskii}}\ and\ \bibinfo {author} {\bibfnamefont {Y.~V.}\ \bibnamefont
  {Nazarov}},\ }\href {\doibase 10.1103/PhysRevB.64.125316} {\bibfield
  {journal} {\bibinfo  {journal} {Physical Review B}\ }\textbf {\bibinfo
  {volume} {64}},\ \bibinfo {pages} {125316} (\bibinfo {year}
  {2001})}\BibitemShut {NoStop}%
\bibitem [{\citenamefont {Woods}\ \emph {et~al.}(2002)\citenamefont {Woods},
  \citenamefont {Reinecke},\ and\ \citenamefont {Lyanda-Geller}}]{Woods2002}%
  \BibitemOpen
  \bibfield  {author} {\bibinfo {author} {\bibfnamefont {L.}~\bibnamefont
  {Woods}}, \bibinfo {author} {\bibfnamefont {T.}~\bibnamefont {Reinecke}}, \
  and\ \bibinfo {author} {\bibfnamefont {Y.}~\bibnamefont {Lyanda-Geller}},\
  }\href {\doibase 10.1103/PhysRevB.69.125330} {\bibfield  {journal} {\bibinfo
  {journal} {Physical Review B}\ }\textbf {\bibinfo {volume} {66}},\ \bibinfo
  {pages} {161318(R)} (\bibinfo {year} {2002})}\BibitemShut {NoStop}%
\bibitem [{\citenamefont {Golovach}\ \emph {et~al.}(2004)\citenamefont
  {Golovach}, \citenamefont {Khaetskii},\ and\ \citenamefont
  {Loss}}]{Golovach2004}%
  \BibitemOpen
  \bibfield  {author} {\bibinfo {author} {\bibfnamefont {V.~N.}\ \bibnamefont
  {Golovach}}, \bibinfo {author} {\bibfnamefont {A.}~\bibnamefont {Khaetskii}},
  \ and\ \bibinfo {author} {\bibfnamefont {D.}~\bibnamefont {Loss}},\ }\href
  {\doibase 10.1103/PhysRevLett.93.016601} {\bibfield  {journal} {\bibinfo
  {journal} {Physical Review Letters}\ }\textbf {\bibinfo {volume} {93}},\
  \bibinfo {pages} {016601} (\bibinfo {year} {2004})}\BibitemShut {NoStop}%
\bibitem [{\citenamefont {Warburton}(2013)}]{Warburton2013}%
  \BibitemOpen
  \bibfield  {author} {\bibinfo {author} {\bibfnamefont {R.~J.}\ \bibnamefont
  {Warburton}},\ }\href {\doibase 10.1038/nmat3585} {\bibfield  {journal}
  {\bibinfo  {journal} {Nature Materials}\ }\textbf {\bibinfo {volume} {12}},\
  \bibinfo {pages} {483} (\bibinfo {year} {2013})}\BibitemShut {NoStop}%
\bibitem [{\citenamefont {Prechtel}\ \emph {et~al.}(2015)\citenamefont
  {Prechtel}, \citenamefont {Kuhlmann}, \citenamefont {Houel}, \citenamefont
  {Ludwig}, \citenamefont {Valentin}, \citenamefont {Wieck},\ and\
  \citenamefont {Warburton}}]{Prechtel2015}%
  \BibitemOpen
  \bibfield  {author} {\bibinfo {author} {\bibfnamefont {J.~H.}\ \bibnamefont
  {Prechtel}}, \bibinfo {author} {\bibfnamefont {A.~V.}\ \bibnamefont
  {Kuhlmann}}, \bibinfo {author} {\bibfnamefont {J.}~\bibnamefont {Houel}},
  \bibinfo {author} {\bibfnamefont {A.}~\bibnamefont {Ludwig}}, \bibinfo
  {author} {\bibfnamefont {S.~R.}\ \bibnamefont {Valentin}}, \bibinfo {author}
  {\bibfnamefont {A.~D.}\ \bibnamefont {Wieck}}, \ and\ \bibinfo {author}
  {\bibfnamefont {R.~J.}\ \bibnamefont {Warburton}},\ }\href@noop {} {\bibfield
   {journal} {\bibinfo  {journal} {Unpublished}\ } (\bibinfo {year}
  {2015})}\BibitemShut {NoStop}%
\bibitem [{\citenamefont {Rice}\ and\ \citenamefont
  {Carmichael}(1988)}]{Rice1988}%
  \BibitemOpen
  \bibfield  {author} {\bibinfo {author} {\bibfnamefont {P.~R.}\ \bibnamefont
  {Rice}}\ and\ \bibinfo {author} {\bibfnamefont {H.~J.}\ \bibnamefont
  {Carmichael}},\ }\href {\doibase 10.1109/3.974} {\bibfield  {journal}
  {\bibinfo  {journal} {IEEE Journal of Quantum Electronics}\ }\textbf
  {\bibinfo {volume} {24}},\ \bibinfo {pages} {1351} (\bibinfo {year}
  {1988})}\BibitemShut {NoStop}%
\bibitem [{\citenamefont {Rephaeli}\ and\ \citenamefont
  {Fan}(2013)}]{Rephaeli2013}%
  \BibitemOpen
  \bibfield  {author} {\bibinfo {author} {\bibfnamefont {E.}~\bibnamefont
  {Rephaeli}}\ and\ \bibinfo {author} {\bibfnamefont {S.}~\bibnamefont {Fan}},\
  }\href {\doibase 10.1364/PRJ.1.000110} {\bibfield  {journal} {\bibinfo
  {journal} {Photonics Research}\ }\textbf {\bibinfo {volume} {1}},\ \bibinfo
  {pages} {110} (\bibinfo {year} {2013})}\BibitemShut {NoStop}%
\bibitem [{\citenamefont {Ralph}\ \emph {et~al.}(2015)\citenamefont {Ralph},
  \citenamefont {S{\"{o}}llner}, \citenamefont {Mahmoodian}, \citenamefont
  {White},\ and\ \citenamefont {Lodahl}}]{Ralph2015}%
  \BibitemOpen
  \bibfield  {author} {\bibinfo {author} {\bibfnamefont {T.}~\bibnamefont
  {Ralph}}, \bibinfo {author} {\bibfnamefont {I.}~\bibnamefont
  {S{\"{o}}llner}}, \bibinfo {author} {\bibfnamefont {S.}~\bibnamefont
  {Mahmoodian}}, \bibinfo {author} {\bibfnamefont {A.}~\bibnamefont {White}}, \
  and\ \bibinfo {author} {\bibfnamefont {P.}~\bibnamefont {Lodahl}},\ }\href
  {\doibase 10.1103/PhysRevLett.114.173603} {\bibfield  {journal} {\bibinfo
  {journal} {Physical Review Letters}\ }\textbf {\bibinfo {volume} {114}},\
  \bibinfo {pages} {173603} (\bibinfo {year} {2015})}\BibitemShut {NoStop}%
\bibitem [{\citenamefont {Bradford}\ \emph {et~al.}(2012)\citenamefont
  {Bradford}, \citenamefont {Obi},\ and\ \citenamefont {Shen}}]{Bradford2012}%
  \BibitemOpen
  \bibfield  {author} {\bibinfo {author} {\bibfnamefont {M.}~\bibnamefont
  {Bradford}}, \bibinfo {author} {\bibfnamefont {K.~C.}\ \bibnamefont {Obi}}, \
  and\ \bibinfo {author} {\bibfnamefont {J.~T.}\ \bibnamefont {Shen}},\ }\href
  {\doibase 10.1103/PhysRevLett.108.103902} {\bibfield  {journal} {\bibinfo
  {journal} {Physical Review Letters}\ }\textbf {\bibinfo {volume} {108}},\
  \bibinfo {pages} {103902} (\bibinfo {year} {2012})}\BibitemShut {NoStop}%
\bibitem [{\citenamefont {Shomroni}\ \emph {et~al.}(2014)\citenamefont
  {Shomroni}, \citenamefont {Rosenblum}, \citenamefont {Lovsky}, \citenamefont
  {Bechler}, \citenamefont {Guendelman},\ and\ \citenamefont
  {Dayan}}]{Shomroni2014}%
  \BibitemOpen
  \bibfield  {author} {\bibinfo {author} {\bibfnamefont {I.}~\bibnamefont
  {Shomroni}}, \bibinfo {author} {\bibfnamefont {S.}~\bibnamefont {Rosenblum}},
  \bibinfo {author} {\bibfnamefont {Y.}~\bibnamefont {Lovsky}}, \bibinfo
  {author} {\bibfnamefont {O.}~\bibnamefont {Bechler}}, \bibinfo {author}
  {\bibfnamefont {G.}~\bibnamefont {Guendelman}}, \ and\ \bibinfo {author}
  {\bibfnamefont {B.}~\bibnamefont {Dayan}},\ }\href {\doibase
  10.1126/science.1254699} {\bibfield  {journal} {\bibinfo  {journal}
  {Science}\ }\textbf {\bibinfo {volume} {345}},\ \bibinfo {pages} {903}
  (\bibinfo {year} {2014})}\BibitemShut {NoStop}%
\bibitem [{\citenamefont {Arcari}\ \emph {et~al.}(2014)\citenamefont {Arcari},
  \citenamefont {S{\"{o}}llner}, \citenamefont {Javadi}, \citenamefont
  {Hansen}, \citenamefont {Mahmoodian}, \citenamefont {Liu}, \citenamefont
  {Thyrrestrup}, \citenamefont {Lee}, \citenamefont {Song}, \citenamefont
  {Stobbe},\ and\ \citenamefont {Lodahl}}]{Arcari2014}%
  \BibitemOpen
  \bibfield  {author} {\bibinfo {author} {\bibfnamefont {M.}~\bibnamefont
  {Arcari}}, \bibinfo {author} {\bibfnamefont {I.}~\bibnamefont
  {S{\"{o}}llner}}, \bibinfo {author} {\bibfnamefont {A.}~\bibnamefont
  {Javadi}}, \bibinfo {author} {\bibfnamefont {S.~L.}\ \bibnamefont {Hansen}},
  \bibinfo {author} {\bibfnamefont {S.}~\bibnamefont {Mahmoodian}}, \bibinfo
  {author} {\bibfnamefont {J.}~\bibnamefont {Liu}}, \bibinfo {author}
  {\bibfnamefont {H.}~\bibnamefont {Thyrrestrup}}, \bibinfo {author}
  {\bibfnamefont {E.~H.}\ \bibnamefont {Lee}}, \bibinfo {author} {\bibfnamefont
  {J.~D.}\ \bibnamefont {Song}}, \bibinfo {author} {\bibfnamefont
  {S.}~\bibnamefont {Stobbe}}, \ and\ \bibinfo {author} {\bibfnamefont
  {P.}~\bibnamefont {Lodahl}},\ }\href {\doibase
  10.1103/PhysRevLett.113.093603} {\bibfield  {journal} {\bibinfo  {journal}
  {Physical Review Letters}\ }\textbf {\bibinfo {volume} {13}},\ \bibinfo
  {pages} {093603} (\bibinfo {year} {2014})}\BibitemShut {NoStop}%
\bibitem [{\citenamefont {Wang}\ \emph {et~al.}(2011)\citenamefont {Wang},
  \citenamefont {Stobbe},\ and\ \citenamefont {Lodahl}}]{Wang2011}%
  \BibitemOpen
  \bibfield  {author} {\bibinfo {author} {\bibfnamefont {Q.}~\bibnamefont
  {Wang}}, \bibinfo {author} {\bibfnamefont {S.}~\bibnamefont {Stobbe}}, \ and\
  \bibinfo {author} {\bibfnamefont {P.}~\bibnamefont {Lodahl}},\ }\href
  {\doibase 10.1103/PhysRevLett.107.167404} {\bibfield  {journal} {\bibinfo
  {journal} {Physical Review Letters}\ }\textbf {\bibinfo {volume} {107}},\
  \bibinfo {pages} {167404} (\bibinfo {year} {2011})}\BibitemShut {NoStop}%
\bibitem [{Note2()}]{Note2}%
  \BibitemOpen
  \bibinfo {note} {\label {footnote2}See Supplemental Material at [URL will be
  inserted by publisher] for more information on the reduction of the crystal
  symmetry.}\BibitemShut {Stop}%
\bibitem [{\citenamefont {Wen}\ \emph {et~al.}(2008)\citenamefont {Wen},
  \citenamefont {David}, \citenamefont {Checoury}, \citenamefont {{El Kurdi}},\
  and\ \citenamefont {Boucaud}}]{Wen2008}%
  \BibitemOpen
  \bibfield  {author} {\bibinfo {author} {\bibfnamefont {F.}~\bibnamefont
  {Wen}}, \bibinfo {author} {\bibfnamefont {S.}~\bibnamefont {David}}, \bibinfo
  {author} {\bibfnamefont {X.}~\bibnamefont {Checoury}}, \bibinfo {author}
  {\bibfnamefont {M.}~\bibnamefont {{El Kurdi}}}, \ and\ \bibinfo {author}
  {\bibfnamefont {P.}~\bibnamefont {Boucaud}},\ }\href {\doibase
  10.1364/OE.16.012278} {\bibfield  {journal} {\bibinfo  {journal} {Optics
  Express}\ }\textbf {\bibinfo {volume} {16}},\ \bibinfo {pages} {12278}
  (\bibinfo {year} {2008})}\BibitemShut {NoStop}%
\bibitem [{\citenamefont {Trif}\ \emph {et~al.}(2009)\citenamefont {Trif},
  \citenamefont {Simon},\ and\ \citenamefont {Loss}}]{Trif2009}%
  \BibitemOpen
  \bibfield  {author} {\bibinfo {author} {\bibfnamefont {M.}~\bibnamefont
  {Trif}}, \bibinfo {author} {\bibfnamefont {P.}~\bibnamefont {Simon}}, \ and\
  \bibinfo {author} {\bibfnamefont {D.}~\bibnamefont {Loss}},\ }\href {\doibase
  10.1103/PhysRevLett.103.106601} {\bibfield  {journal} {\bibinfo  {journal}
  {Physical Review Letters}\ }\textbf {\bibinfo {volume} {103}},\ \bibinfo
  {pages} {106601} (\bibinfo {year} {2009})}\BibitemShut {NoStop}%
\bibitem [{\citenamefont {Rogers}\ \emph {et~al.}(2014)\citenamefont {Rogers},
  \citenamefont {Jahnke}, \citenamefont {Metsch}, \citenamefont {Sipahigil},
  \citenamefont {Binder}, \citenamefont {Teraji}, \citenamefont {Sumiya},
  \citenamefont {Isoya}, \citenamefont {Lukin}, \citenamefont {Hemmer},\ and\
  \citenamefont {Jelezko}}]{Rogers2014b}%
  \BibitemOpen
  \bibfield  {author} {\bibinfo {author} {\bibfnamefont {L.~J.}\ \bibnamefont
  {Rogers}}, \bibinfo {author} {\bibfnamefont {K.~D.}\ \bibnamefont {Jahnke}},
  \bibinfo {author} {\bibfnamefont {M.~H.}\ \bibnamefont {Metsch}}, \bibinfo
  {author} {\bibfnamefont {A.}~\bibnamefont {Sipahigil}}, \bibinfo {author}
  {\bibfnamefont {J.~M.}\ \bibnamefont {Binder}}, \bibinfo {author}
  {\bibfnamefont {T.}~\bibnamefont {Teraji}}, \bibinfo {author} {\bibfnamefont
  {H.}~\bibnamefont {Sumiya}}, \bibinfo {author} {\bibfnamefont
  {J.}~\bibnamefont {Isoya}}, \bibinfo {author} {\bibfnamefont {M.~D.}\
  \bibnamefont {Lukin}}, \bibinfo {author} {\bibfnamefont {P.}~\bibnamefont
  {Hemmer}}, \ and\ \bibinfo {author} {\bibfnamefont {F.}~\bibnamefont
  {Jelezko}},\ }\href {\doibase 10.1103/PhysRevLett.113.263602} {\bibfield
  {journal} {\bibinfo  {journal} {Physical Review Letters}\ }\textbf {\bibinfo
  {volume} {113}},\ \bibinfo {pages} {263602} (\bibinfo {year}
  {2014})}\BibitemShut {NoStop}%
\bibitem [{\citenamefont {Pingault}\ \emph {et~al.}(2014)\citenamefont
  {Pingault}, \citenamefont {Becker}, \citenamefont {Schulte}, \citenamefont
  {Arend}, \citenamefont {Hepp}, \citenamefont {Godde}, \citenamefont
  {Tartakovskii}, \citenamefont {Markham}, \citenamefont {Becher},\ and\
  \citenamefont {Atat{\"{u}}re}}]{Pingault2014}%
  \BibitemOpen
  \bibfield  {author} {\bibinfo {author} {\bibfnamefont {B.}~\bibnamefont
  {Pingault}}, \bibinfo {author} {\bibfnamefont {J.~N.}\ \bibnamefont
  {Becker}}, \bibinfo {author} {\bibfnamefont {C.~H.}\ \bibnamefont {Schulte}},
  \bibinfo {author} {\bibfnamefont {C.}~\bibnamefont {Arend}}, \bibinfo
  {author} {\bibfnamefont {C.}~\bibnamefont {Hepp}}, \bibinfo {author}
  {\bibfnamefont {T.}~\bibnamefont {Godde}}, \bibinfo {author} {\bibfnamefont
  {A.~I.}\ \bibnamefont {Tartakovskii}}, \bibinfo {author} {\bibfnamefont
  {M.}~\bibnamefont {Markham}}, \bibinfo {author} {\bibfnamefont
  {C.}~\bibnamefont {Becher}}, \ and\ \bibinfo {author} {\bibfnamefont
  {M.}~\bibnamefont {Atat{\"{u}}re}},\ }\href {\doibase
  10.1103/PhysRevLett.113.263601} {\bibfield  {journal} {\bibinfo  {journal}
  {Physical Review Letters}\ }\textbf {\bibinfo {volume} {113}},\ \bibinfo
  {pages} {263601} (\bibinfo {year} {2014})}\BibitemShut {NoStop}%
\bibitem [{\citenamefont {Davan{\c{c}}o}\ \emph {et~al.}(2012)\citenamefont
  {Davan{\c{c}}o}, \citenamefont {Chan}, \citenamefont {Safavi-Naeini},
  \citenamefont {Painter},\ and\ \citenamefont {Srinivasan}}]{Davanco2012}%
  \BibitemOpen
  \bibfield  {author} {\bibinfo {author} {\bibfnamefont {M.}~\bibnamefont
  {Davan{\c{c}}o}}, \bibinfo {author} {\bibfnamefont {J.}~\bibnamefont {Chan}},
  \bibinfo {author} {\bibfnamefont {A.~H.}\ \bibnamefont {Safavi-Naeini}},
  \bibinfo {author} {\bibfnamefont {O.}~\bibnamefont {Painter}}, \ and\
  \bibinfo {author} {\bibfnamefont {K.}~\bibnamefont {Srinivasan}},\ }\href
  {\doibase 10.1364/OE.20.024394} {\bibfield  {journal} {\bibinfo  {journal}
  {Optics Express}\ }\textbf {\bibinfo {volume} {20}},\ \bibinfo {pages}
  {24394} (\bibinfo {year} {2012})}\BibitemShut {NoStop}%
\bibitem [{\citenamefont {Bochmann}\ \emph {et~al.}(2013)\citenamefont
  {Bochmann}, \citenamefont {Vainsencher}, \citenamefont {Awschalom},\ and\
  \citenamefont {Cleland}}]{Bochmann2013}%
  \BibitemOpen
  \bibfield  {author} {\bibinfo {author} {\bibfnamefont {J.}~\bibnamefont
  {Bochmann}}, \bibinfo {author} {\bibfnamefont {A.}~\bibnamefont
  {Vainsencher}}, \bibinfo {author} {\bibfnamefont {D.~D.}\ \bibnamefont
  {Awschalom}}, \ and\ \bibinfo {author} {\bibfnamefont {A.~N.}\ \bibnamefont
  {Cleland}},\ }\href {\doibase 10.1038/nphys2748} {\bibfield  {journal}
  {\bibinfo  {journal} {Nature Physics}\ }\textbf {\bibinfo {volume} {9}},\
  \bibinfo {pages} {712} (\bibinfo {year} {2013})}\BibitemShut {NoStop}%
\bibitem [{\citenamefont {Duan}\ \emph {et~al.}(2001)\citenamefont {Duan},
  \citenamefont {Lukin}, \citenamefont {Cirac},\ and\ \citenamefont
  {Zoller}}]{Duan2001}%
  \BibitemOpen
  \bibfield  {author} {\bibinfo {author} {\bibfnamefont {L.~M.}\ \bibnamefont
  {Duan}}, \bibinfo {author} {\bibfnamefont {M.~D.}\ \bibnamefont {Lukin}},
  \bibinfo {author} {\bibfnamefont {J.~I.}\ \bibnamefont {Cirac}}, \ and\
  \bibinfo {author} {\bibfnamefont {P.}~\bibnamefont {Zoller}},\ }\href
  {\doibase 10.1038/35106500} {\bibfield  {journal} {\bibinfo  {journal}
  {Nature}\ }\textbf {\bibinfo {volume} {414}},\ \bibinfo {pages} {413}
  (\bibinfo {year} {2001})}\BibitemShut {NoStop}%
\bibitem [{\citenamefont {Kepesidis}\ \emph {et~al.}(2013)\citenamefont
  {Kepesidis}, \citenamefont {Bennett}, \citenamefont {Portolan}, \citenamefont
  {Lukin},\ and\ \citenamefont {Rabl}}]{Kepesidis2013}%
  \BibitemOpen
  \bibfield  {author} {\bibinfo {author} {\bibfnamefont {K.~V.}\ \bibnamefont
  {Kepesidis}}, \bibinfo {author} {\bibfnamefont {S.~D.}\ \bibnamefont
  {Bennett}}, \bibinfo {author} {\bibfnamefont {S.}~\bibnamefont {Portolan}},
  \bibinfo {author} {\bibfnamefont {M.~D.}\ \bibnamefont {Lukin}}, \ and\
  \bibinfo {author} {\bibfnamefont {P.}~\bibnamefont {Rabl}},\ }\href {\doibase
  10.1103/PhysRevB.88.064105} {\bibfield  {journal} {\bibinfo  {journal}
  {Physical Review B}\ }\textbf {\bibinfo {volume} {88}},\ \bibinfo {pages}
  {064105} (\bibinfo {year} {2013})}\BibitemShut {NoStop}%
\bibitem [{\citenamefont {S{\"{o}}llner}\ \emph {et~al.}(2015)\citenamefont
  {S{\"{o}}llner}, \citenamefont {Mahmoodian}, \citenamefont {Hansen},
  \citenamefont {Midolo}, \citenamefont {Javadi}, \citenamefont
  {Kir{\v{s}}ansk{\.{e}}}, \citenamefont {Pregnolato}, \citenamefont {El-Ella},
  \citenamefont {Lee}, \citenamefont {Song}, \citenamefont {Stobbe},\ and\
  \citenamefont {Lodahl}}]{Sollner2015}%
  \BibitemOpen
  \bibfield  {author} {\bibinfo {author} {\bibfnamefont {I.}~\bibnamefont
  {S{\"{o}}llner}}, \bibinfo {author} {\bibfnamefont {S.}~\bibnamefont
  {Mahmoodian}}, \bibinfo {author} {\bibfnamefont {S.~L.}\ \bibnamefont
  {Hansen}}, \bibinfo {author} {\bibfnamefont {L.}~\bibnamefont {Midolo}},
  \bibinfo {author} {\bibfnamefont {A.}~\bibnamefont {Javadi}}, \bibinfo
  {author} {\bibfnamefont {G.}~\bibnamefont {Kir{\v{s}}ansk{\.{e}}}}, \bibinfo
  {author} {\bibfnamefont {T.}~\bibnamefont {Pregnolato}}, \bibinfo {author}
  {\bibfnamefont {H.}~\bibnamefont {El-Ella}}, \bibinfo {author} {\bibfnamefont
  {E.~H.}\ \bibnamefont {Lee}}, \bibinfo {author} {\bibfnamefont {J.~D.}\
  \bibnamefont {Song}}, \bibinfo {author} {\bibfnamefont {S.}~\bibnamefont
  {Stobbe}}, \ and\ \bibinfo {author} {\bibfnamefont {P.}~\bibnamefont
  {Lodahl}},\ }\href {\doibase 10.1038/nnano.2015.159} {\bibfield  {journal}
  {\bibinfo  {journal} {Nature Nanotechnology}\ }\textbf {\bibinfo {volume}
  {10}},\ \bibinfo {pages} {775} (\bibinfo {year} {2015})}\BibitemShut
  {NoStop}%
\end{thebibliography}

\begin{thebibliography}{8}%
\makeatletter
\providecommand \@ifxundefined [1]{%
 \@ifx{#1\undefined}
}%
\providecommand \@ifnum [1]{%
 \ifnum #1\expandafter \@firstoftwo
 \else \expandafter \@secondoftwo
 \fi
}%
\providecommand \@ifx [1]{%
 \ifx #1\expandafter \@firstoftwo
 \else \expandafter \@secondoftwo
 \fi
}%
\providecommand \natexlab [1]{#1}%
\providecommand \enquote  [1]{``#1''}%
\providecommand \bibnamefont  [1]{#1}%
\providecommand \bibfnamefont [1]{#1}%
\providecommand \citenamefont [1]{#1}%
\providecommand \href@noop [0]{\@secondoftwo}%
\providecommand \href [0]{\begingroup \@sanitize@url \@href}%
\providecommand \@href[1]{\@@startlink{#1}\@@href}%
\providecommand \@@href[1]{\endgroup#1\@@endlink}%
\providecommand \@sanitize@url [0]{\catcode `\\12\catcode `\$12\catcode
  `\&12\catcode `\#12\catcode `\^12\catcode `\_12\catcode `\%12\relax}%
\providecommand \@@startlink[1]{}%
\providecommand \@@endlink[0]{}%
\providecommand \url  [0]{\begingroup\@sanitize@url \@url }%
\providecommand \@url [1]{\endgroup\@href {#1}{\urlprefix }}%
\providecommand \urlprefix  [0]{URL }%
\providecommand \Eprint [0]{\href }%
\providecommand \doibase [0]{http://dx.doi.org/}%
\providecommand \selectlanguage [0]{\@gobble}%
\providecommand \bibinfo  [0]{\@secondoftwo}%
\providecommand \bibfield  [0]{\@secondoftwo}%
\providecommand \translation [1]{[#1]}%
\providecommand \BibitemOpen [0]{}%
\providecommand \bibitemStop [0]{}%
\providecommand \bibitemNoStop [0]{.\EOS\space}%
\providecommand \EOS [0]{\spacefactor3000\relax}%
\providecommand \BibitemShut  [1]{\csname bibitem#1\endcsname}%
\let\auto@bib@innerbib\@empty
%</preamble>
\bibitem [{\citenamefont {Sakoda}(2005)}]{Sakoda2005}%
  \BibitemOpen
  \bibfield  {author} {\bibinfo {author} {\bibfnamefont {K.}~\bibnamefont
  {Sakoda}},\ }\href@noop {} {\emph {\bibinfo {title} {Optical Properties of
  Photonic Crystals}}},\ \bibinfo {edition} {2nd}\ ed.,\ edited by\ \bibinfo
  {editor} {\bibfnamefont {W.}~\bibnamefont {Rhodes}}\ (\bibinfo  {publisher}
  {Springer Berlin Heidelberg New York},\ \bibinfo {year} {2005})\ p.\ \bibinfo
  {pages} {253}\BibitemShut {NoStop}%
\bibitem [{\citenamefont {Mirman}(1999)}]{Mirman99}%
  \BibitemOpen
  \bibfield  {author} {\bibinfo {author} {\bibfnamefont {R.}~\bibnamefont
  {Mirman}},\ }\href {https://books.google.nl/books?id=owhE36qiTP8C} {\emph
  {\bibinfo {title} {{Point Groups, Space Groups, Crystals, Molecules}}}}\
  (\bibinfo  {publisher} {World Scientific},\ \bibinfo {year}
  {1999})\BibitemShut {NoStop}%
\bibitem [{\citenamefont {Safavi-Naeini}\ and\ \citenamefont
  {Painter}(2010)}]{Safavi-Naeini2010aS}%
  \BibitemOpen
  \bibfield  {author} {\bibinfo {author} {\bibfnamefont {A.~H.}\ \bibnamefont
  {Safavi-Naeini}}\ and\ \bibinfo {author} {\bibfnamefont {O.}~\bibnamefont
  {Painter}},\ }\href {\doibase 10.1364/OE.18.014926} {\bibfield  {journal}
  {\bibinfo  {journal} {Optics Express}\ }\textbf {\bibinfo {volume} {18}},\
  \bibinfo {pages} {14926} (\bibinfo {year} {2010})}\BibitemShut {NoStop}%
\bibitem [{\citenamefont {Terzibaschian}\ and\ \citenamefont
  {Enderlein}(1986)}]{Terzibaschian1986}%
  \BibitemOpen
  \bibfield  {author} {\bibinfo {author} {\bibfnamefont {T.}~\bibnamefont
  {Terzibaschian}}\ and\ \bibinfo {author} {\bibfnamefont {R.}~\bibnamefont
  {Enderlein}},\ }\href {\doibase 10.1002/pssb.2221330202} {\bibfield
  {journal} {\bibinfo  {journal} {Physica Status Solidi (B)}\ }\textbf
  {\bibinfo {volume} {133}},\ \bibinfo {pages} {443} (\bibinfo {year}
  {1986})}\BibitemShut {NoStop}%
\bibitem [{\citenamefont {Andreani}\ and\ \citenamefont
  {Gerace}(2006)}]{Andreani2006S}%
  \BibitemOpen
  \bibfield  {author} {\bibinfo {author} {\bibfnamefont {L.~C.}\ \bibnamefont
  {Andreani}}\ and\ \bibinfo {author} {\bibfnamefont {D.}~\bibnamefont
  {Gerace}},\ }\href {\doibase 10.1103/PhysRevB.73.235114} {\bibfield
  {journal} {\bibinfo  {journal} {Physical Review B}\ }\textbf {\bibinfo
  {volume} {73}},\ \bibinfo {pages} {235114} (\bibinfo {year}
  {2006})}\BibitemShut {NoStop}%
\bibitem [{\citenamefont {Wen}\ \emph {et~al.}(2008)\citenamefont {Wen},
  \citenamefont {David}, \citenamefont {Checoury}, \citenamefont {{El Kurdi}},\
  and\ \citenamefont {Boucaud}}]{Wen2008S}%
  \BibitemOpen
  \bibfield  {author} {\bibinfo {author} {\bibfnamefont {F.}~\bibnamefont
  {Wen}}, \bibinfo {author} {\bibfnamefont {S.}~\bibnamefont {David}}, \bibinfo
  {author} {\bibfnamefont {X.}~\bibnamefont {Checoury}}, \bibinfo {author}
  {\bibfnamefont {M.}~\bibnamefont {{El Kurdi}}}, \ and\ \bibinfo {author}
  {\bibfnamefont {P.}~\bibnamefont {Boucaud}},\ }\href {\doibase
  10.1364/OE.16.012278} {\bibfield  {journal} {\bibinfo  {journal} {Optics
  Express}\ }\textbf {\bibinfo {volume} {16}},\ \bibinfo {pages} {12278}
  (\bibinfo {year} {2008})}\BibitemShut {NoStop}%
\bibitem [{\citenamefont {Johnson}\ and\ \citenamefont
  {Joannopoulos}(2001)}]{Johnson2001}%
  \BibitemOpen
  \bibfield  {author} {\bibinfo {author} {\bibfnamefont {S.}~\bibnamefont
  {Johnson}}\ and\ \bibinfo {author} {\bibfnamefont {J.}~\bibnamefont
  {Joannopoulos}},\ }\href {\doibase 10.1364/OE.8.000173} {\bibfield  {journal}
  {\bibinfo  {journal} {Optics Express}\ }\textbf {\bibinfo {volume} {8}},\
  \bibinfo {pages} {173} (\bibinfo {year} {2001})}\BibitemShut {NoStop}%
\bibitem [{Note3()}]{Note3}%
  \BibitemOpen
  \bibinfo {note} {Can be found at: \par
  http://www.ioffe.ru/SVA/NSM/Semicond/GaAs/mechanic.html.}\BibitemShut {Stop}%
\end{thebibliography}

\section{Supplementary Material for ``Deterministic Single-Phonon Source Triggered by a Single Photon''}
\subsection{The Irreducible Brillouin Zone of the Shamrock-crystal}

\begin{figure*} [ht!]
	\includegraphics{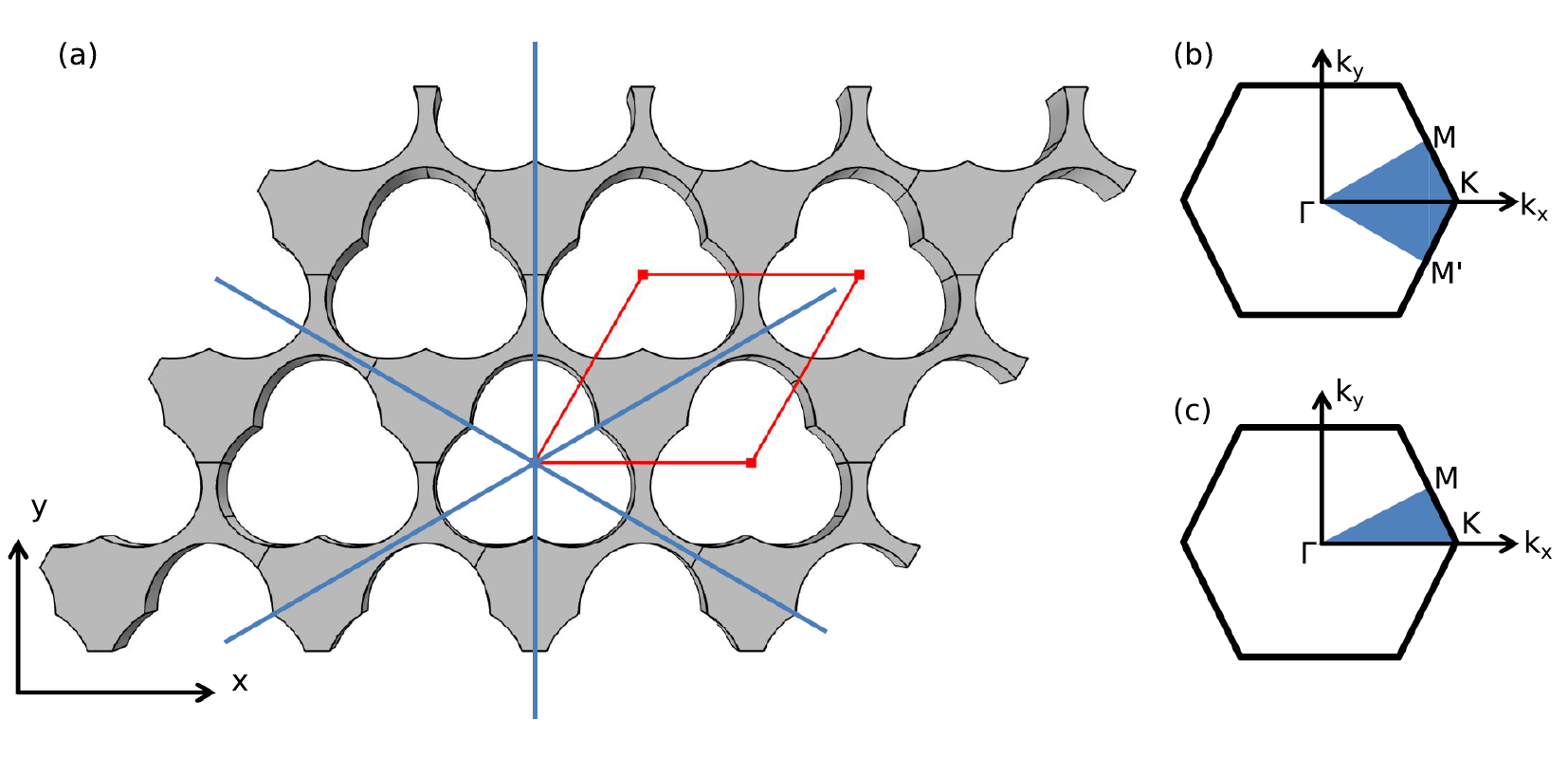}
\caption{ \label{fig:Sfig1} The symmetry of the shamrock-crystal. {\bf a)} Three solid blue lines indicate the mirror planes of the $\textrm{C}_{3v}$ symmetry of the shamrock-hole. The red rhombus indicates the primitive cell of the crystal. The particular orientation bewteen the mirror planes an the vectors forming the primitive basis leads to a crystal structure that belongs to the p3m1 planar space group. Thus, distinctly different then the p6mm planar space group, which can be obtained for holes that have at least as high a degree of symmetry as the hexagonal lattice. {\bf b)} In general both the photonic and the phononic eigenmodes inherit the symmetries of the crystal-structure and hence the irreducible Brillouin zone can be found from the planar space group of the structure. Here the irreducible Brillouin zone for p3m1 is shown \cite{Terzibaschian1986}. However, this planar space group does not include the inversion operator while the optical and the mechanical modes must both be eigenstates of the inversion operator, due to time-reversal symmetry ($\omega(k)=\omega(-k)$) \cite{Andreani2006S, Wen2008S}, leading to a smaller effective irreducible Brillouin zone. {\bf c)} The effective irreducible Brillouin zone for the optical and mechanical modes in a crystal belonging to the p3m1 planar space group. }
\end{figure*}

Most work to date on photonic-crystals uses circularly symmetric holes ($\textrm{C}_{\infty}$), typically in a hexagonal lattice ($\textrm{C}_{6v}$) \cite{Sakoda2005}, leading to a crystal that belongs to the planar space-group (also called wallpaper group) p6mm \cite{Mirman99}. Recent studies investigating novel opto-mechanical crystals, known as snowflake-crystals \cite{Safavi-Naeini2010aS}, the symmetry of the hole has been reduced to match the symmetry of the lattice ($\textrm{C}_{6v}$). This again leads to a crystal that belongs to the planar space-group p6mm. However, for the shamrock-crystal, c.f. Fig.~\ref{fig:Sfig1}a), the symmetry of the hole is reduced further by removing three of the mirror planes which results in a $\textrm{C}_{3v}$ symmetry of the hole. For the orientation of the hole with respect to the lattice shown in Fig.~\ref{fig:Sfig1}a) the total structure belongs to the planar space-group p3m1 \cite{Mirman99}. The symmetries of the photonic and phononic modes are inherited from the symmetries of the structure as is the irreducible Brillouin zone (IRBZ), c.f. Fig.~\ref{fig:Sfig1}b)\cite{Terzibaschian1986}. In addition, the photonic and phononic modes must be eigenstates of the inversion operator, which is not one of the symmetry operators of the planar space-group p3m1. This arises because all modes fulfilling time-reversal symmetry must be eigenstates of the inversion operator. Since inversion was not a symmetry operator of the initial crystal structure this changes the IRBZ shown in Fig.~\ref{fig:Sfig1}b) to an effective IRBZ, Fig.~\ref{fig:Sfig1}c) \cite{Andreani2006S, Wen2008S}. This is the IRBZ of a crystal that belongs to the standard planar space-group p6mm \cite{Terzibaschian1986} and is used for all simulations presented in this work.

The photonic-bands were calculated using the freely available software package MIT Photonic Bands \cite{Johnson2001}. For the calculations of the photonic modes a refractive index of 3.48 is used, which corresponds to GaAs at $\SI{4}{\kelvin}$. The photonic cavity modes were calculated in a commercially available finite-element solver. This software was also used for the calculation of the phononic-bands and the phononic cavity modes. The large degree of anisotropy in GaAs is taken into account in the phononic simulations, where we have used the three elastic constants relevant for GaAs \footnote{Can be found at:

http://www.ioffe.ru/SVA/NSM/Semicond/GaAs/mechanic.html.}.

\end{document}